\documentclass[aps,twocolumn,showpacs,lengthcheck,preprintnumbers]{revtex4-1}
\usepackage{amsfonts}
\usepackage{amsmath}
\usepackage{amssymb}
\usepackage{graphicx}
\usepackage{fontenc}
\usepackage{color}
\usepackage[unicode]{hyperref}
\hypersetup{
	pdftitle={collisions},
	pdfauthor={},
	pdfsubject={solitons},
	colorlinks=true,
	linkcolor=blue,
	citecolor=blue,
	filecolor=black,
	urlcolor=blue
}

\begin{document}
\title{Soliton collisions in Bose-Einstein condensates with current-dependent interactions }
	
\author{Qian Jia}
\affiliation{
	School of Science, Xi’an University of Posts and Telecommunications, Xi’an, China
}	
\author{Haibo Qiu}
\affiliation{
	School of Science, Xi’an University of Posts and Telecommunications, Xi’an, China
}
\author{A. Mu\~{n}oz Mateo}
\affiliation{Departamento de F\'isica, Universidad de La Laguna, La Laguna, Tenerife, Spain}


\begin{abstract} 
We study general collisions between chiral solitons in Bose-Einstein condensates subject to combined attractive and current-dependent interatomic interactions.
A simple analysis based on the linear superposition of the solitons allows us to determine the
relevant time and space scales of the dynamics, which is illustrated by extensive numerical simulations.
By varying the differential amplitude, the relative phase, the average velocity, and the relative velocity of the solitons,
we characterize the different dynamical regimes that give rise to oscillatory and interference phenomena.
 Apart from the known inelastic character of the collisions, we show that the 
chiral dynamics involves an amplitude reduction with respect to the case of regular solitons.
To compare with feasible ultracold gas experiments, the influence of harmonic confinement is analyzed in both the emergence and the interaction of chiral solitons.

\end{abstract}

\maketitle

\section{Introduction}
\label{sec:intro}
The research on matter-wave solitons entered a new stage since the first experiments on collapsing Bose-Einstein condensates (BECs) of ultracold gases \cite{Sackett1999,Gerton2000}. The whole process of emergence and evolution of bright solitons could be observed in experiments that, by making use of magnetic Feshbach resonances of some atomic species, tuned the interatomic forces
from repulsive to attractive interactions  \cite{Donley2001,Strecker2002}. In this way, stable bright matter solitons were generated in elongated 
condensates with quasi one-dimensional (1D) geometries; in most of the cases, an external harmonic potential is necessary to keep the atomic cloud trapped \cite{AlKhawaja2002}.

The realization of matter-soliton trains led naturally to the study of soliton collisions  \cite{Strecker2002,Cornish2006,Nguyen2014}.
This scenario allowed for the experimental test in ultracold gases of predictions that had been made long before for bright soliton interactions in optical fibers \cite{Gordon1983,Desem1987}. In parallel, theoretical studies on matter solitons followed the experimental development \cite{Salasnich2003,Leung2003,Carr2004,Carr2004a,Wuster2009,Billam2012}. 
Currently, the interaction between bright solitons in the framework of the 1D nonlinear Schr\"odinger equation is reasonably well understood as a wave interference process during which Josephson tunneling of particles can take place \cite{Zhao2016,Zhao2017}
However, there are still open, long-standing questions regarding the process of soliton generation and subsequent evolution that need of detailed analysis in order to be settled. To this end, recent experiments that use non-destructive imaging have been carried out in scalar condensates \cite{Everitt2017,Nguyen2017}. 

This year, a new type of matter-wave soliton that shows chiral properties has been observed in experiments with ultracold atoms \cite{Frolian2022}. It was theoretically predicted in 1996 \cite{Aglietti1996}, and
its existence relies on the action of a density-dependent gauge field, which provides the system with chiral properties. The experimental realization of density-dependent gauge fields in ultracold atoms had been achieved in the presence of optical lattices \cite{Clark2018,Gorg2019}, but only very recently it has been realized in translational invariant settings \cite{Yao2022,Frolian2022}. The emergent chiral properties of the system are reflected also in the free expansion of the atomic cloud and the onset of persistent currents \cite{Edmonds2013}, or the center of mass oscillations \cite{Edmonds2015}, and are particularly manifest in the direction-dependent motion (and existence) of bright, chiral solitons \cite{Aglietti1996,Frolian2022}.

 Before the experiment \cite{Frolian2022} took place, chiral solitons had been demonstrated to be dynamically stable objects \cite{Dingwall2019}.
 The collisions between chiral solitons with equal number of particles had been studied \cite{Dingwall2018}, where a non-integrable dynamics stands out as the main difference with respect to the collisions of regular solitons. The action of the modulational instability has also been analyzed in the presence of a density-dependent gauge field and absence of trapping \cite{Bhat2021}, showing the chiral features of the resulting soliton train. Still, as can be inferred from the comparison with the extensive literature on regular solitons, the study of chiral solitons is just starting and requires further characterization, more so with the prospect of experimental test. 
 
  The present paper contributes to this characterization by analyzing general collisions between chiral solitons with different number of particles, including the variation of both the relative phase and the relative velocity.
  The collisions are studied first in the absence of confinement, and later, motivated by the usual experimental settings, within a harmonic trap.
  The soliton emergence is also addressed in order to show the influence of the harmonic confinement.
  We characterize the dynamical regimes of chiral soliton collisions, which are dominated by oscillatory and interference phenomena. The relative phase plays a more decisive role than in regular solitons, since it can determine the transmission and reflection coefficients of the soliton scattering. 
  Our analysis is made in the framework of a generalized Gross-Pitaevskii equation that, besides the usual contact-interaction term, includes a current-dependent interaction as derived from a non-local unitary transformation of the theory containing the density-dependent gauge field  \cite{Aglietti1996}.  

The rest of the paper is structured as follows: Section \ref{sec:Model} makes a detailed introduction of the
system model including the properties of relevant states, plane waves and solitons, trapped and untrapped, and their connection through dynamical decay. Section \ref{sec:Collisions} presents the theoretical basis that rules the soliton collisions and their dynamical regimes, which are tested first for regular solitons, and later for chiral solitons. Section \ref{sec:Conclusions} summarizes our results. The Appendix \ref{sec:units} and \ref{sec:currentInterf} try, respectively, to clarify on the particular units employed in our analysis, and to provide additional details on several aspects of chiral-soliton collisions.

\section{Model}
\label{sec:Model}
We assume that the system is an elongated BEC at zero temperature with frozen
transverse degrees of freedom, such that the order parameter of the three-dimensional (3D) condensate is space separable $\Psi(\mathbf{r},t)=\psi(x,t)\chi(y,z)$, where 
$\chi(y,z)$ is the transverse ground state. 
We further assume, within a mean field framework, that the axial wave function $\psi(x,t)$ follows a generalized 1D Gross-Pitaevskii equation 
\begin{equation}
i\hbar\frac{\partial\psi}{\partial t}=\left[ 
-\frac{\hbar^{2}}{2m}\frac{\partial^2}{\partial 
	x^2}+U_{\rm ext}(x)+g_{\mbox{\tiny 1D}}\left\vert \psi\right\vert ^{2} +\hbar\kappa J\right]  \psi,
\label{eq:gpe}
\end{equation}
where  $U_{\rm ext}(x)$ is an external axial potential,  $g_{\mbox{\tiny 1D}}<0$ is the strength of the usual contact interparticle interaction, and $J(x,t)$ is the current density  $J=\hbar(\psi^*\partial_x\psi-\psi\partial_x\psi^*)/(i2m)$. The latter quantity, which introduces a current-dependent mean field, can be demonstrated to enter the equation of motion, in a different representation (see Ref. \cite{Aglietti1996} for details), through a density-dependent gauge field that induces a momentum shift of value $\hbar\kappa |\psi|^2/2$, where $\kappa$ is dimensionless.
 
After multiplication on the left of Eq. (\ref{eq:gpe}) by $\psi^*$, and subtracting the resulting equation from its complex conjugate, one obtains the continuity equation
\begin{equation}
\frac{\partial}{\partial t}|\psi|^2+\frac{\partial}{\partial x} J =0,
\label{eq:continuity}
\end{equation}
which, from the integration over the whole space $d/dt\int\,dx\,|\psi|^2=0$, gives the conservation of the number of particles $N=\int dx|\psi|^2$.
Additionally, the Hamiltonian operator in Eq. (\ref{eq:gpe}), $H_{\mbox{\tiny GP}}=\hat p^2/2m+U_{\rm ext}+g_{\mbox{\tiny 1D}}|\psi|^2+\hbar \kappa J$, endows the system with unusual global properties. The expectation value of the momentum operator $\langle\hat p\rangle=-i\hbar\int dx\,\psi^*\partial_x\psi$, which follows the generic equation $i\hbar{d}/{d t}\langle\hat p\rangle= \langle [\hat p,\,H_{\mbox{\tiny GP}}]\rangle$,  gives, making use of the continuity equation,
\begin{equation}
\frac{d}{d t}\left\langle\hat p-\frac{\hbar\kappa}{2}|\psi|^2\right\rangle= \left\langle -\frac{\partial U_{\rm ext}}{\partial x}\right\rangle.
\label{eq:momentum}
\end{equation}
In the absence of external potential, the total mechanical momentum $\Pi=\int dx\,\psi^*(\hat p-{\hbar\kappa}|\psi|^2/2)\psi$ is conserved. Analogously, the expectation value of the Hamiltonian,
which follows an equation for a time-dependent operator, gives ${d}/{d t}\langle H_{\mbox{\tiny GP}}\rangle= \langle \partial_t(g_{\mbox{\tiny 1D}}|\psi|^2+\hbar\kappa J)\rangle$, and by using again the continuity equation, ${d}/{d t}\langle H_{\mbox{\tiny GP}}\rangle={d}/{d t}(\langle g_{\mbox{\tiny 1D}}|\psi|^2/2+\hbar\kappa J\rangle)$, that is
\begin{equation}
\frac{d}{d t}\int dx\, \psi^*\left( -\frac{\hbar^{2}}{2m}\frac{\partial^2}{\partial 
	x^2}+U_{\rm ext}+\frac{g_{\mbox{\tiny 1D}}}{2}\left\vert \psi\right\vert ^{2} \right)\psi=0,
\label{eq:energy}
\end{equation}
so, the total energy $E$ given by the above, $\kappa$-independent integral, is a conserved quantity \cite{Aglietti1996}. 

It is insightful to rewrite the
current density as $J=|\psi|^2\,v$, where the superfluid velocity $v(x,t)=\hbar\partial_x S/m$ is defined from the wave function phase $S(x,t)=\arg\,\psi$; hence, Eq. (\ref{eq:gpe}) can be recast as
\begin{equation}
i\hbar\frac{\partial\psi}{\partial t}=\left[
-\frac{\hbar^{2}}{2m}\frac{\partial^2}{\partial 
	x^2}+U_{\rm ext}+g_v\left\vert \psi\right\vert ^{2} \right]  \psi,
\label{eq:gpk}
\end{equation}
where the velocity-dependent effective interaction $g_v$ is defined by 
\begin{align}
g_v(x,t)=g_{\mbox{\tiny 1D}}+\hbar\kappa\, v(x,t) .
\label{eq:gv}
\end{align}
Therefore, the effective interparticle interaction $g_v$ changes its character (hence its sign) from attractive to repulsive when the local velocity exceeds the limit value set by the contact interaction $ v>(v_g\equiv |g_{\mbox{\tiny 1D}}/\hbar\kappa|)$, otherwise the effective interaction remains attractive for  $v<v_g$.

The stationary states present the coordinate separable wave function $\psi(x,t)=\psi(x)\,\exp(-i\mu t/\hbar)$, where $\mu$ is the eigenvalue of the Hamiltonian operator $H_{\mbox{\tiny GP}}\psi=\mu\psi$, but differently to the regular Gross-Pitaevskii equation, $\mu$ is not (in general) the chemical potential $\mu_{\rm ch}=\partial_N E$. The spectrum of linear excitations $\delta \psi_j= [u_j,\,v_j]^T$ of stationary states, so that $\psi(x,t)\rightarrow \exp(-i\mu t/\hbar)\,\{\psi(x)+\sum_j [u_j(x)\,\exp(-i\omega_j t)+v_j(x)^*\,\exp(i\omega_j t)]\}$, with $j$ being a mode index, can be obtained through the 2$\times$2 Bogoliubov equations $B\delta\psi_j=\hbar\omega_j\,\delta\psi_j$, where the Bogoliubov matrix can be written as $B=B_{\mbox{\tiny GP}}+B_\kappa$, explicitly,
\begin{align}
B_{\mbox{\tiny GP}}=
\begin{pmatrix}
H_{\mbox{\tiny GP}}+g_{\mbox{\tiny 1D}}|\psi|^2-\mu & g_{\mbox{\tiny 1D}}\psi^2\\
- g_{\mbox{\tiny 1D}}{\psi^*}^2& -H_{\mbox{\tiny GP}}-g_{\mbox{\tiny 1D}}|\psi|^2+\mu
\end{pmatrix},
\label{eq:Bog1}
\end{align}
and
\begin{align}
B_\kappa=i\frac{\hbar^2\kappa}{2m}
\begin{pmatrix}
\psi{\partial_x \psi^*}-|\psi|^2{\partial_x} & -\psi\partial_x \psi+\psi^2{\partial_x}\\
 -\psi^*{\partial_x \psi^*}+{\psi^*}^2{\partial_x}& \psi^*{\partial_x \psi}-|\psi|^2{\partial_x}
\end{pmatrix}.
\label{eq:Bog2}
\end{align}
The existence of complex frequencies in the spectrum of linear excitations, that is $\Im(\omega_j)\neq 0$, indicates the presence of unstable modes that have an exponential growth (in the linear regime) from perturbative values, thus capable of breaking the stationary configuration.

\subsection{Plane waves}
For $U_{\rm ext}=0$, Eq. (\ref{eq:gpe}) is translational invariant; in this case, it is useful to look at the spectrum of plane wave eigenstates $\psi_q(x,t)=\sqrt{n}\,\exp[i(qx-\mu_q t/\hbar)]$, having shifted frequencies $\omega_q=\mu_q/\hbar$ defined by

\begin{align}
 \mu_q=\hbar^2q^2/2m+(g_{\mbox{\tiny 1D}}+\hbar^2\kappa q/m)\,n.
 \label{eq:muPW}
\end{align}
 
  Therefore, the group velocity
of the waves $v_q=\partial_q \mu_q/\hbar=\hbar(q+\kappa n)/m$ does not match the superfluid velocity $v=\hbar q/m$. 
\begin{figure}[tb]
	\includegraphics[width=\linewidth]{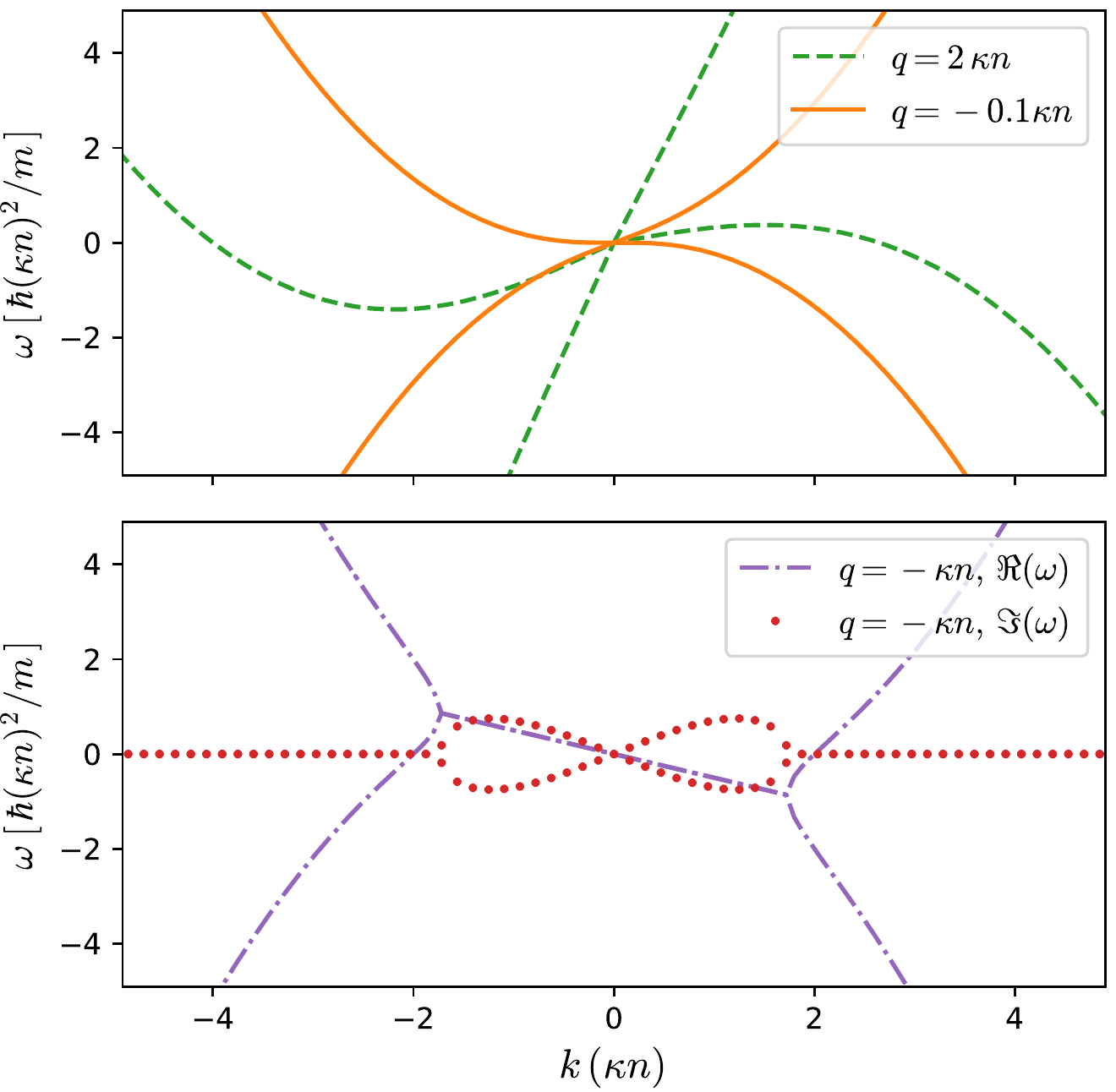}
	\caption{Linear excitation frequencies $\omega$ of plane-wave states with wave number $q$ in a system with current-dependent interactions (and $U_{\rm ext}=0$, $g_{\mbox{\tiny 1D}}=0$). For $q>-\kappa n/4$ (top panel) the plane-wave states are dynamically stable (all the excitation modes present real frequencies), and the speed of sound, $c_q^\pm=\partial_k \omega|_{k\rightarrow 0}$, is different for rightward and leftward moving waves.  For $q<-\kappa n/4 $ the plane wave states are unstable against long wave length perturbation modes (bottom panel), which have imaginary frequencies.}
	\label{fig:bog_planew}
\end{figure}

The linear excitations of a plane wave $\psi_q$ can also be expanded in Fourier modes $\delta\psi_k=e^{ikx}\{u\,\exp(iqx),\,v\,\exp(-iqx)\}^T$ that produce independent, algebraic Bogoliubov equations for each excitation mode with wave vector $k$. The resulting excitation
dispersion is
\begin{align}
\omega_{k\pm}=\frac{\hbar k}{m}\left(q+\frac{\kappa n}{2}\pm\sqrt{\frac{k^2+(\kappa n)^2}{4}+q\,\kappa n +\frac{m\,g_{\mbox{\tiny 1D}}n}{\hbar^2} }\right)
\label{eq:planew_disp}
\end{align}
which is asymmetric, $|\omega_{k+}|\neq |\omega_{k-}|$ (see the top panel of Fig. \ref{fig:bog_planew}), even as looked at from a moving frame with velocity $\hbar q/m$, and only becomes symmetric if looked at from a reference frame moving with velocity $\hbar (q+\kappa n/2)/m$. This fact reflects the origin of the current-density term in the generalized GP Eq. (\ref{eq:gpe}), which involves a momentum shift of $\hbar\kappa n/2$ due to the action of a density-dependent gauge field \cite{Aglietti1996}.
From the dispersion Eq. (\ref{eq:planew_disp}), the speed of sound $c_\pm=\partial_k \omega_{k\pm}$ is obtained for long wave length excitations $k\rightarrow 0$ as
\begin{align}
c_\pm= \frac{\hbar }{m}\left(q+\frac{\kappa n}{2}\pm \sqrt{\frac{(\kappa n)^2}{4}+q\,\kappa n +\frac{m\,g_{\mbox{\tiny 1D}}n}{\hbar^2} }\right),
\label{eq:sound}
\end{align}
which also shows the chiral features ($|c_+|\neq |c_-|$) of the system. 

\begin{figure}[tb]
	\centering
	\flushleft \textbf{(a)} 
	\includegraphics[width=0.97\linewidth]{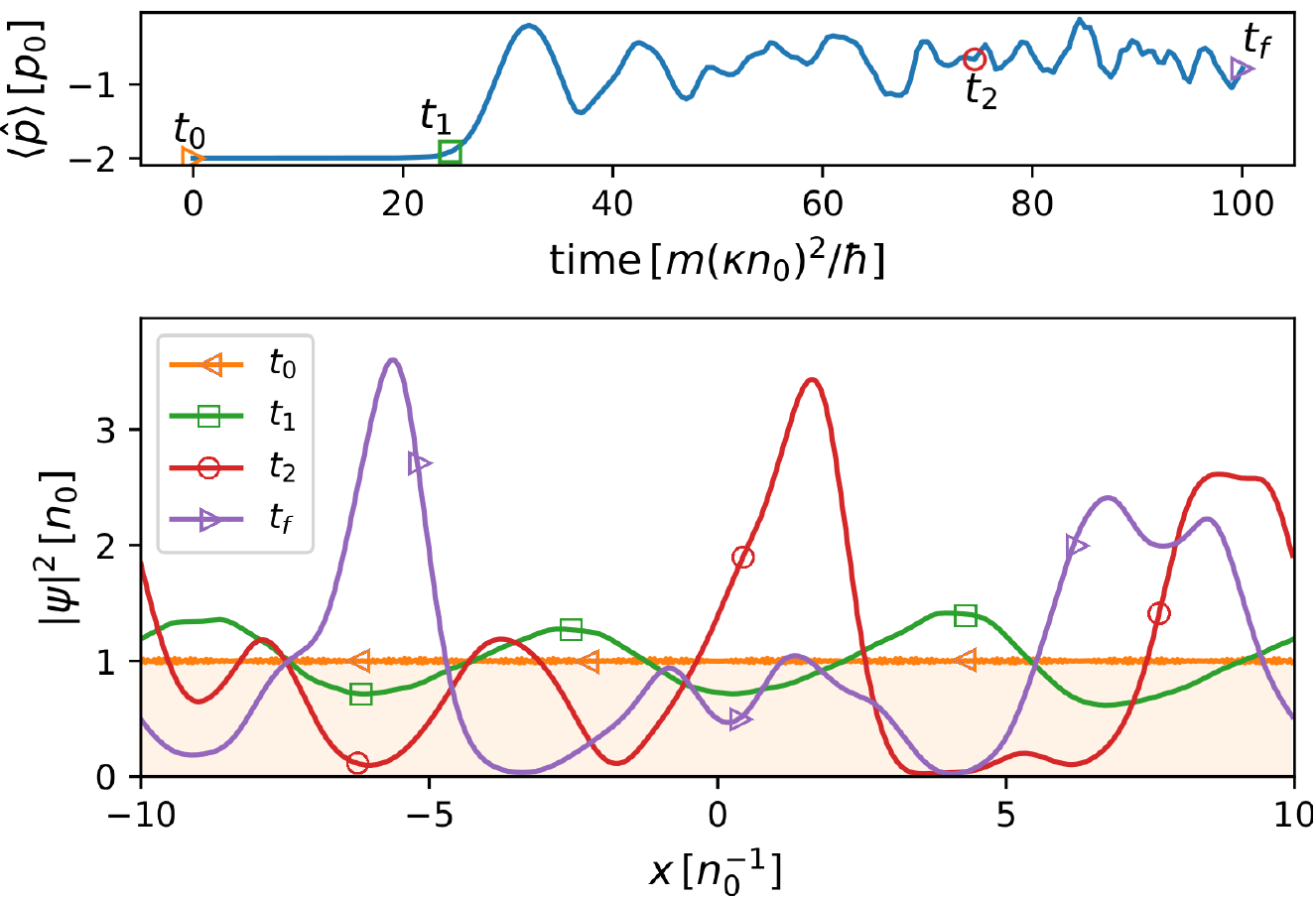}\\
	\vspace{-0.5cm}
	\flushleft \textbf{(b)}
	\includegraphics[width=\linewidth]{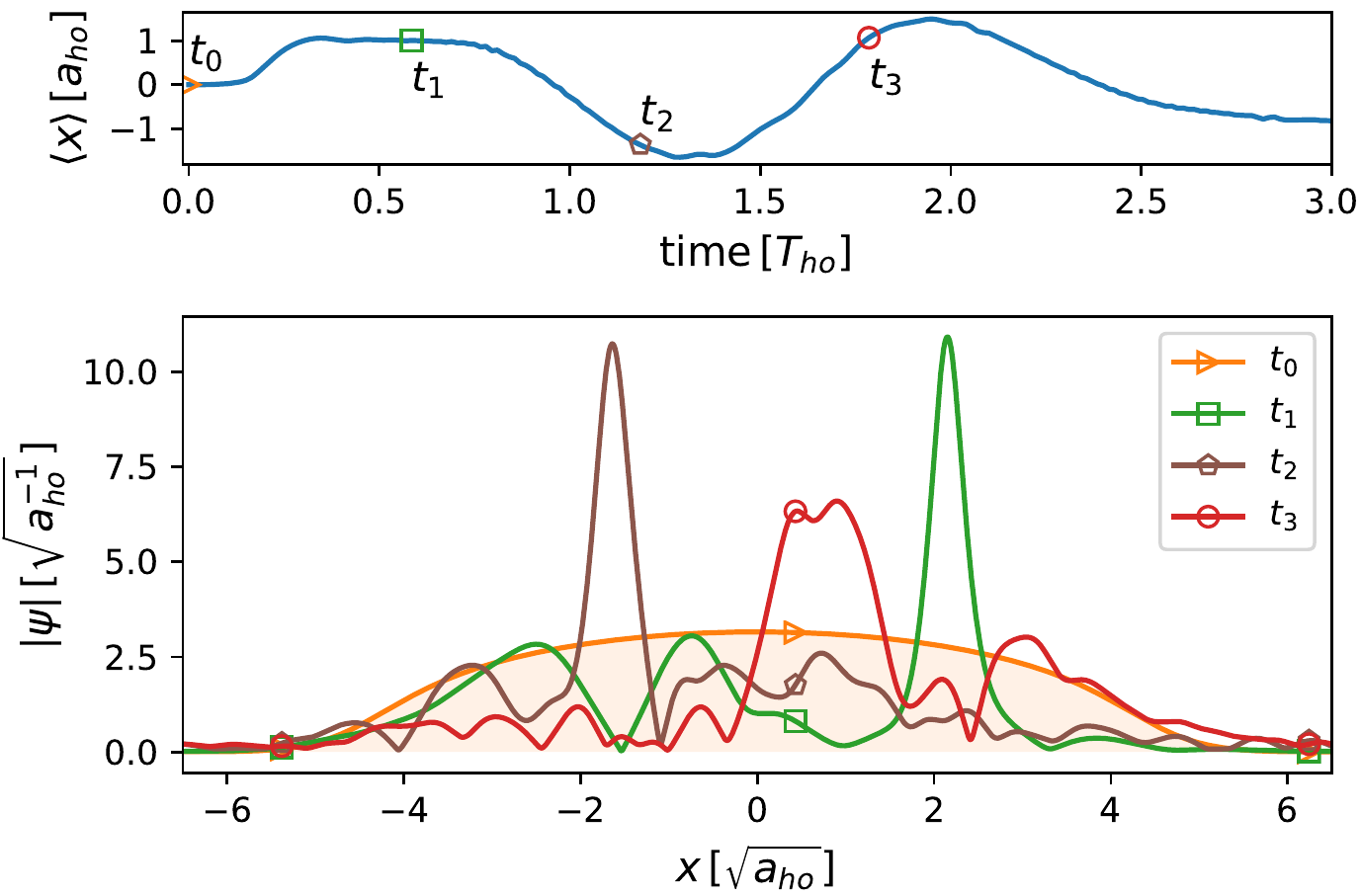}
	
	\caption{(a) Decay of a plane wave state $\psi(t_0)\equiv\psi_q=\sqrt{n_0}\exp(-i4\hbar \pi\,x/L)$, seeded with a random perturbation, in a finite domain of length $L=20\,n_0^{-1}$ with periodic boundary conditions. The interactions are set by $g_{\mathrm{\tiny 1D}}=0$ and  $\kappa=1$. The expectation value of the canonical momentum $\langle\hat p\rangle$ as a function of time (top panel), and snapshots of the density profile at selected times (bottom panel) are shown.  (b) Decay in a harmonic trap after a sudden change of the interatomic interactions from repulsive at $t=t_0$, with $g_{\mathrm{\tiny 1D}}=1\,\hbar\omega_{\rm ho} a_{\rm ho}$ and $\kappa=0$, to attractive for $t>t_0$, with $g_{\mathrm{\tiny 1D}}=-0.2\,\hbar\omega_{\rm ho} a_{\rm ho}$ and $\kappa=0.1$. The top panel shows the evolution of the center of mass. The initial wave function $\psi(t_0)$  corresponds to the system ground state when the interactions are repulsive. }
	\label{fig:decay}
\end{figure}
	As follows from Eq. (\ref{eq:planew_disp}), plane waves with wave vector $q<-[\kappa n/4+m\,g_{\mbox{\tiny 1D}}/(\hbar^2\kappa)]$ are unstable against perturbations \cite{Bhat2021}. 
	The bottom panel of Fig. \ref{fig:bog_planew} shows an example with $g_{\mbox{\tiny 1D}}=0$ and $q=-\kappa n$; in this case all wave vectors below $-\kappa n/4$ correspond to unstable states.
	Interestingly, in contrast with the case of just attractive interactions ($\kappa=0$), it is possible to find dynamically stable states with negative effective interaction $g_v=\hbar^2\kappa q/m<0$ in the range $q\in[-\kappa n/4+m\,g_{\mbox{\tiny 1D}}/(\hbar^2\kappa),\,0]$. 
	In particular, for $g_{\mbox{\tiny 1D}}=0$, the set of states with wave vectors  $q\in[-\kappa n/4,\,0]$ are dynamically stable. From the inspection of Eq. (\ref{eq:planew_disp}), one can see that the cause of this extra stability resides in the zero point energy $(\hbar\kappa n)^2/4m$ associated with moving excitation modes. In finite systems, due to the discrete spectrum, the stability window is enlarged by, approximately $(2\pi/L)^2/(4\kappa n)$, where $L$ is the system size, within the domain of negative wave numbers.

	 The dynamical decay of unstable plane waves gives rise to the segmentation of the initial constant density into localized, moving wave packets, akin to bright solitons, that interact with each other \cite{Bhat2021}. This process has been observed in BECs with attractive contact interactions \cite{AlKhawaja2002}, where the resulting number of solitons can be approximated by the ratio $L/\lambda_{max}$, where $\lambda_{max}=2\pi/ k_{max}$ is the wave length of the unstable mode with maximum imaginary frequency $\max[\Im(\omega_k)]$ \cite{Nguyen2017,Bhat2021,Sanz2022}. Apart from the conservation of the total mechanical momentum Eq. (\ref{eq:momentum}) instead of the canonical momentum, an analogous process is followed in the presence of current-dependent interactions (see the recent work \cite{Bhat2021} for details). Figure \ref{fig:decay} illustrates the decay process of an unstable plane wave state in a finite system of size $L=20\,n_0^{-1}$, where $n_0$ is the constant density, in agreement with the predictions of the linear analysis; notice (top panel) that the canonical momentum is not a conserved quantity. The decay is apparent after $t=20\,m(\kappa n_0)^2/\hbar$, as reflected by the wavy density at $t=t_1$, which is consistent with the typical time scale taken for the perturbation growth as set by the maximum imaginary frequency $\max[\Im(\omega_k)]^{-1}=5.4\,\,m(\kappa n_0)^2/\hbar$; moving and interacting soliton-like density peaks are observed afterwards, as for $t=t_2,\,t_f$.

	The emergence of solitons from the decay of a smooth density profile is usually realized under harmonic trapping in ultracold gas experiments  (see for instance Refs. \cite{AlKhawaja2002,Nguyen2017}). In such a setting, the system is subject to a quench in the interatomic interactions, which are changed from repulsive to attractive. The plane wave instability analysis presented before provides just an approximation for the expected unstable modes in the inhomogeneous density profile, by assuming that the maximum density of the trapped system matches the plane wave density. The subsequent dynamics in the trap, in the absence of current-dependent interaction and once the solitons have emerged, follows harmonic cycles of compression and expansion of the whole atomic cloud. As we show in Fig. \ref{fig:decay}(b), the situation is clearly different for $\kappa\neq0$, since the Kohn theorem is not fulfilled \cite{Edmonds2015}, and then the system dynamics does not show harmonic oscillations.

\subsection{Stationary bright solitons}

In the absence of both axial potential and current-dependent interaction, that is $U_{\rm ext}=0$ and $\kappa=0$, Eq.(\ref{eq:gpe}) admits moving bright soliton solutions
\begin{equation}
\psi=\sqrt{\frac{N}{2\xi}} \; \mbox{sech}\left(\frac{x-vt}{\xi}\right)\, 
e^{i[m\,v\,x-(\mu +mv^2/2)t]/\hbar},
\label{eq:sol} 
\end{equation}
where $N$ is the number of particles, $\xi=2\hbar^2 /(m 
|g_{\mbox{\tiny 1D}}| N)$ is the soliton width, and
$\mu=-\hbar^2/(2m\xi^2)=-m g_{\mbox{\tiny 1D}}^2 N^2/8\hbar^3$. Note that the soliton amplitude $A=\sqrt{{N/}{2\xi}}\propto \sqrt{|g_{\mbox{\tiny 1D}}|} N$ is directly proportional to the number of particles.
 Due to the $U(1)$ symmetry of the system, a global, constant phase $\theta_0$ can be added to the soliton phase without affecting observable features such as energy or current.
 For $\kappa=0$ the system is Galilean invariant, so the soliton density profile is independent of the soliton velocity $v$.

\begin{figure}[tb]
	\includegraphics[width=\linewidth]{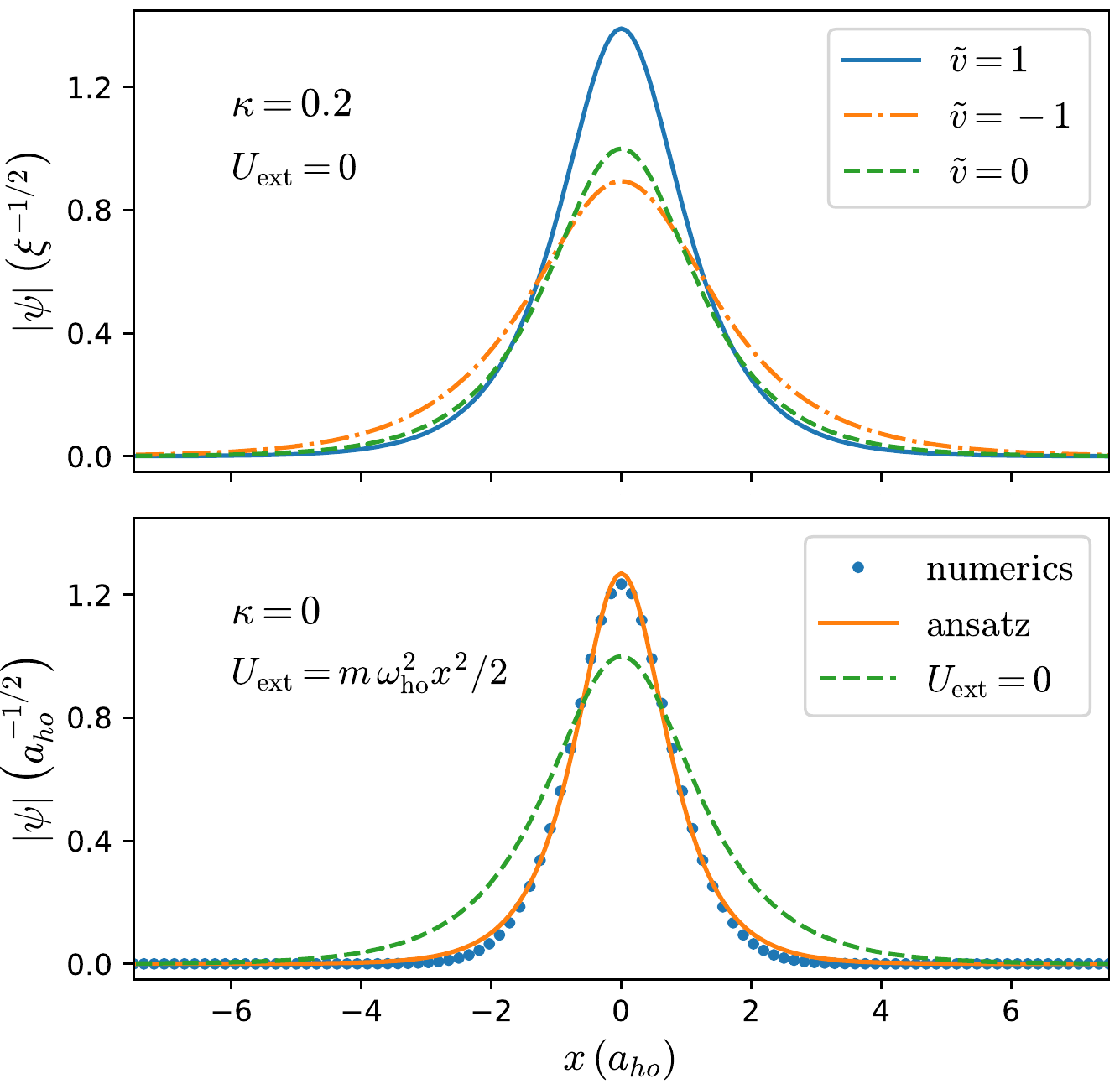}
	\caption{Profile of bright solitons with varying velocity, parameterized by $\tilde v= m v\,\xi /\hbar$, in the presence (top panel) and absence (bottom panel) of current-dependent interactions for fixed number of particles $N=2\times\hbar\omega\,a_{\rm ho}/|g_{\mbox{\tiny 1D}}|$ and contact interaction $g_{\mbox{\tiny 1D}}<0$. In the latter case, the comparison is made for static solitons in the presence and absence of harmonic trap by setting $a_{\rm ho}/\xi=1$, where $\xi=2\hbar^2 /(m |g_{\mbox{\tiny 1D}}| N)$  is the soliton width found in the absence of trap $U_{\rm ext}=0$; see main text for the ansatz description.  }
	\label{fig:widths}
\end{figure}

When $\kappa\neq0$, the bright soliton state Eq. (\ref{eq:sol}) is still a steady wave solution to Eq. (\ref{eq:gpe})  whenever $v<v_g$, however it acquires chiral properties \cite{Aglietti1996}. 
Due to the current-dependent interaction, the soliton width varies with the velocity as $\xi(N)\rightarrow \xi(N,\,v)=2\hbar^2 /(m |g_v| N)$, that is $ \xi(N,\,v)/\xi(N)=|g_{\mbox{\tiny 1D}}/g_v|=(1-v/v_g)^{-1}$. From Eq. (\ref{eq:energy}), the soliton energy is $E_s=mg_v^2N^3(1-|g_{\mbox{\tiny 1D}}/g_v|)/(24\hbar^2)+Nmv^2/2$.
Beyond a velocity threshold $v>v_g$ the bright soliton Eq. (\ref{eq:sol}) is no longer a solution to the GP Eq. (\ref{eq:gpe}). When the stationary soliton exists, it is dynamically stable \cite{Dingwall2019}. The top panel of Fig. \ref{fig:widths} illustrates the different soliton profiles for varying velocity at fixed particle number and contact interaction; for $v=0$ the profile matches the profile of a regular soliton (with $\kappa=0$).

\subsubsection{Harmonic confinement}
\label{sec:harmonic}

Due to its experimental relevance in ultracold gases, we consider also bright soliton states in the presence of harmonic trapping $U_{\rm ext}=m\,\omega_{\rm ho}^2x^2/2$. In this case,
a variational approach provides a good approximation to the exact solutions, see for instance Ref. \cite{Billam2012}. We use the ansatz (in full units) $\psi(x)= \sqrt{N/2\zeta}\,\mbox{sech}(x/\zeta)$, with the width $\zeta(N,\omega)$ as variational parameter for the stationary solution. The minimization of the energy functional, as defined in Eq. (\ref{eq:energy}), $E=\int dx\,[\hbar^2|\partial_x\psi|^2/2m+m\omega_{\rm ho}^2x^2|\psi|^2/2+g_{\mbox{\tiny 1D}}|\psi|^4/2]$, produces a quartic polynomial in $\zeta$ with a single parameter $a_{\rm ho}/\xi$, where $a_{\rm ho}=\sqrt{\hbar/m\omega_{\rm ho}}$ is the trap characteristic length, and  $\xi=2\hbar^2 /(m 
|g_{\mbox{\tiny 1D}}| N)$  is the soliton width found in the absence of trap.
We approximate the solution to the quartic polynomial up to second order in $a_{\rm ho}/\xi$ by $\zeta\, \approx \sqrt{{2}/{\pi}}[1-{2\,a_{\rm ho}/(9\xi)}]\,a_{\rm ho}$    for     $a_{\rm ho}/\xi \in \,\left[ 0,\, {\pi^2}/{4}\right]$, and $\zeta\, \approx \xi\quad$    otherwise. For fixed number of particles, the soliton width is always narrower $\zeta/\xi<1$ in the trapped system, and the chemical potential becomes
\begin{align}
\mu\approx-\frac{\hbar^2}{2m\xi^2}\left(5-2\frac{\xi}{\zeta}\right)\frac{\xi}{3\zeta}.
\label{eq:mutrap}
\end{align}

The bottom panel of Fig. \ref{fig:widths} depicts these features; as can be seen, the analytical ansatz (solid line) provides a good approximation to the exact numerical result (dots); the free soliton (dashed line) is shown for comparison.

\section{Brigh soliton collisions}
\label{sec:Collisions}

The interaction between regular solitons has been explained, by means of the exact two-soliton solutions to the nonlinear Schr\"odinger equation, as a wave interference process during which Josephson tunneling of particles can take place \cite{Zhao2016,Zhao2017}. In what follows, we elaborate on the same idea, without resorting to the exact, complicated analytical solutions, by using simple physical arguments (in the spirit of Ref. \cite{Snyder1997} on optical solitons).

Our analysis of soliton collisions starts with the superposition of two approaching, initially non-overlapping solitons that solve  Eq. (\ref{eq:gpe}) with particle numbers $N_1$ and $N_2$, and relative global phase $\Theta=\theta_1-\theta_2$:
\begin{align}
\psi(x,t=0)= \psi_{A_1,v_1,\,\Theta}(x+x_0) + \psi_{A_2,v_2,\,0}(x-x_0) ,
\label{eq:ini}
\end{align}
where $d_0=2x_0$ is the initial intersoliton distance.
To describe the dynamics, we will make use of average-soliton units (see the Appendix for details), which we will denote by barred symbols; so the length unit $\bar\xi=2\hbar^2 /(m |g_{\mbox{\tiny 1D}}+\hbar\kappa \bar v| \bar N)$ is based on the average number of particles $\bar N=(N_1+N_2)/2$ and average velocity $\bar v=(v_1+v_2)/2$, and the time unit becomes  $\bar \omega^{-1} =m\bar\xi^2/\hbar$.

\begin{figure*}[htp]
	\centering
	\includegraphics[width=.32\textwidth]{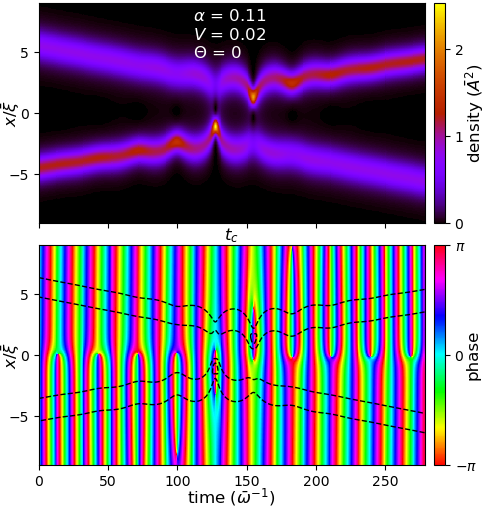}\;
	\includegraphics[width=.32\textwidth]{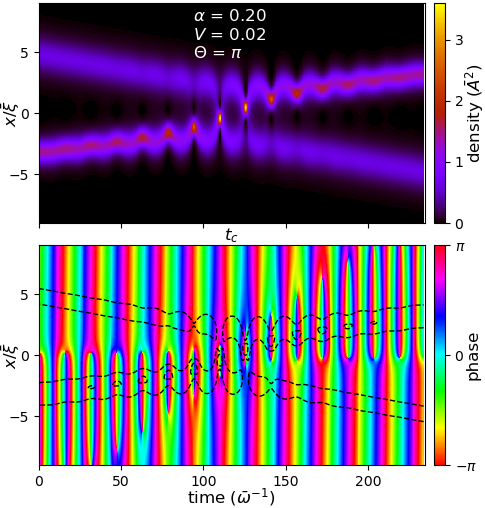}\;
	\includegraphics[width=.32\textwidth]{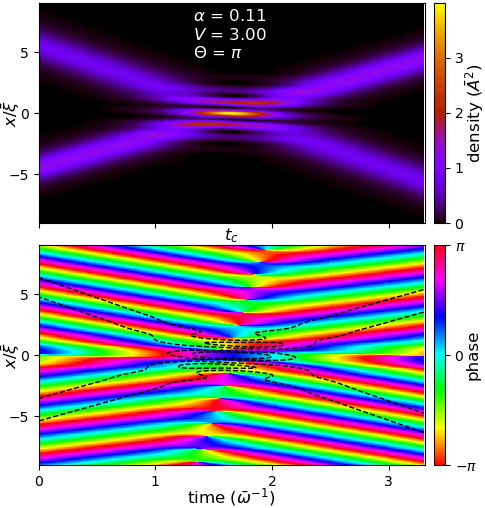}
	\medskip
	\hspace{-0.8cm}
	\includegraphics[width=.29\textwidth]{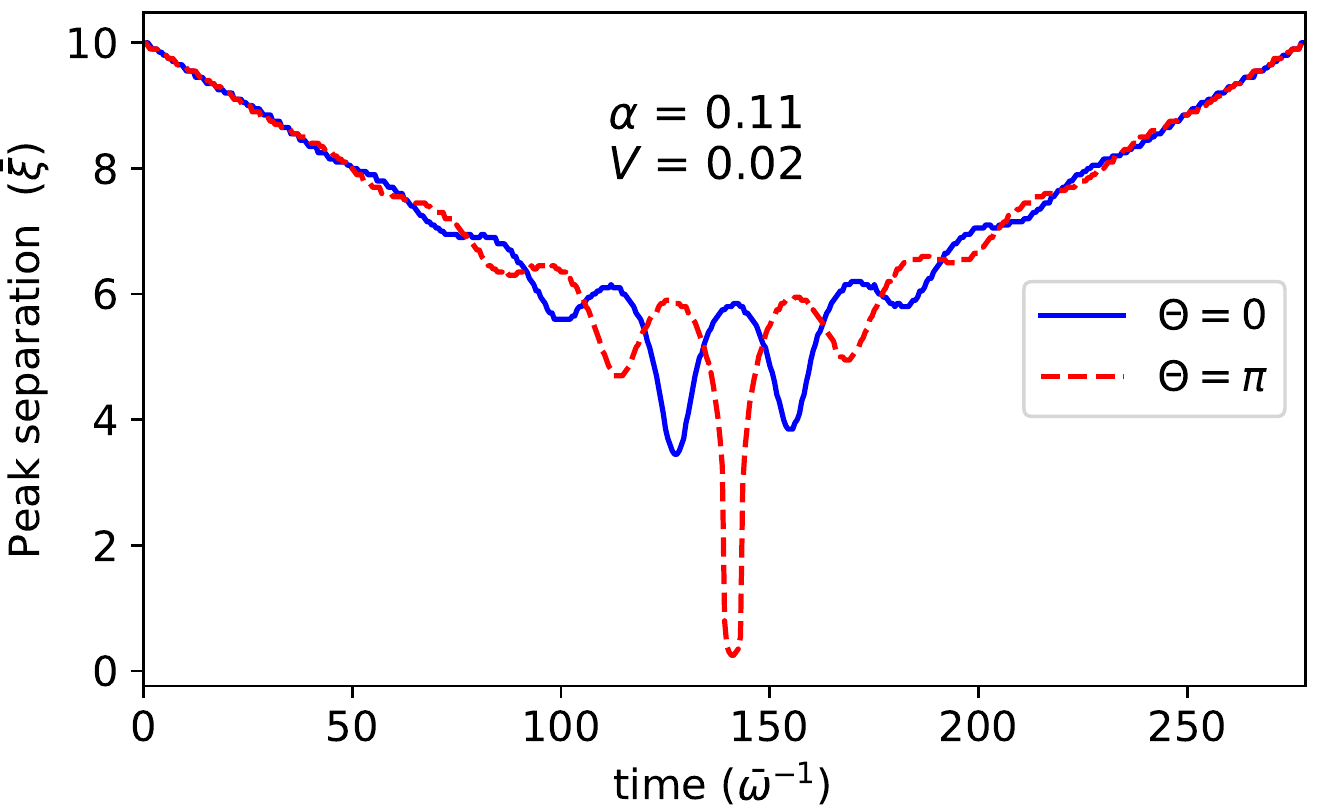}\hspace{0.5cm}	\includegraphics[width=.295\textwidth]{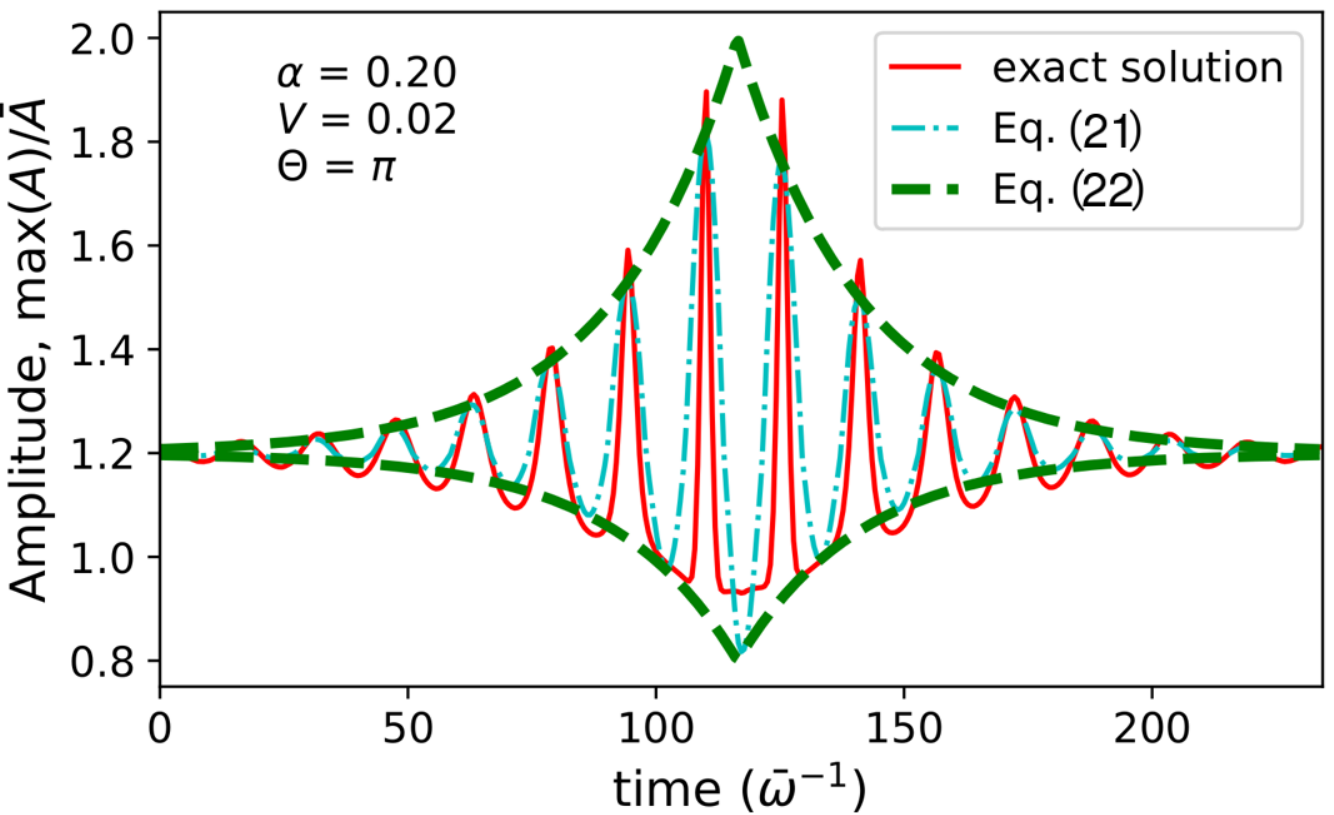}\qquad
	\includegraphics[width=.295\textwidth]{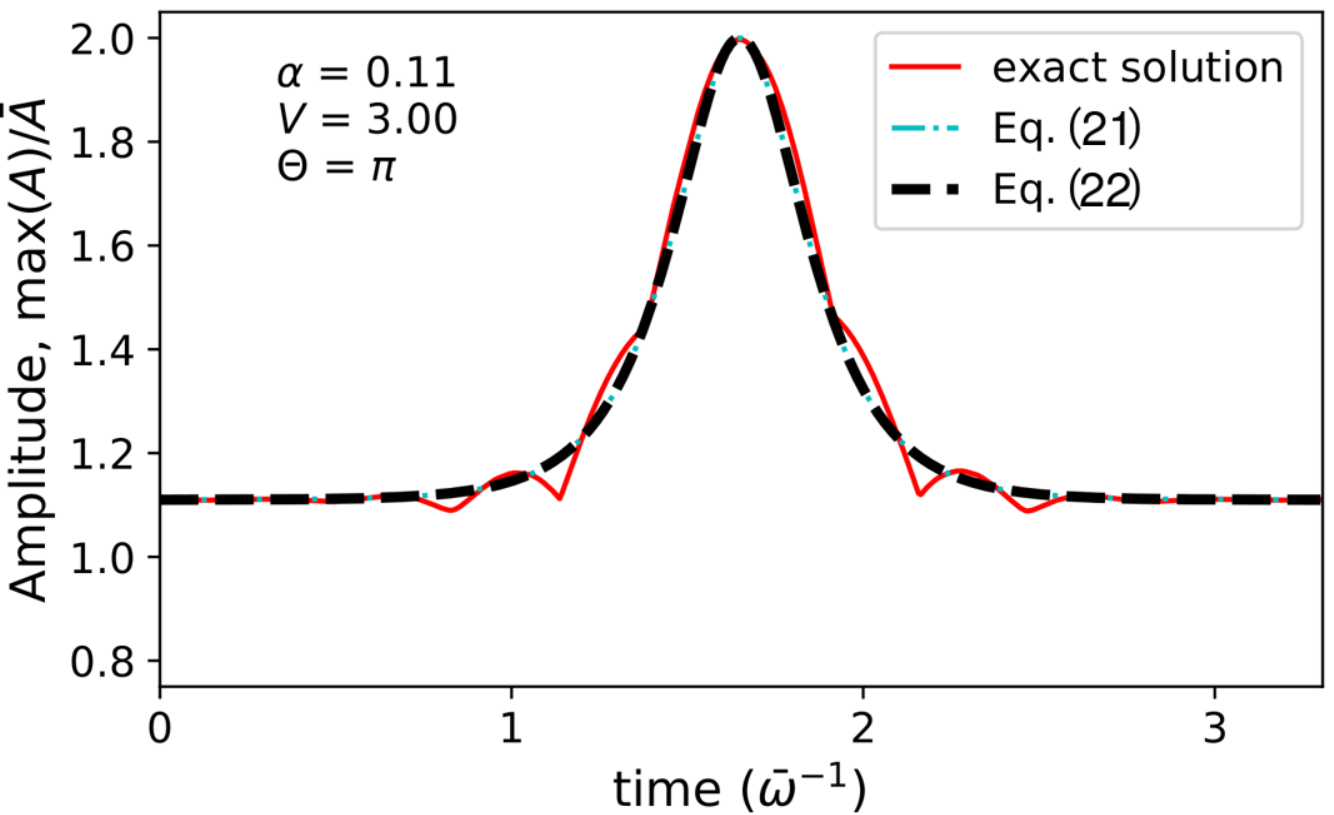}
	
	\caption{Features of soliton collisions within the regular nonlinear Shr\"odinger equation. The time evolution of the density (top panels) and phase profiles (middle panels) are shown for varying collision parameters; the dashed lines in the phase graphs represent two density isocontours. The collision time $t_c$ is indicated on the time axis. Bottom-left panel: soliton separation, as measured between density peaks. Bottom center and right panels:  Time evolution of the maximum density, comparing the exact two-soliton solution with the approximate analytical expression Eqs. (\ref{eq:colleq1}-\ref{eq:colleq2}).  }
	\label{fig:GPcollision}
\end{figure*}

Although the solitons are nonlinear waves, a qualitative picture of the soliton interactions can be obtained from the usual superposition of linear waves. 
Due to the coherent properties of the underlying Bose-Einstein condensate, the non-overlapping solitons of Eq. (\ref{eq:ini}), separated by the distance $d=2(x_0-vt)$, give rise to a neat interference pattern in momentum space of period \cite{Pitaevskii1999}
\begin{align}
	k_d =\frac{\pi}{x_0- vt},
\label{eq:kfringes}
\end{align}
  where we assumed equal relative velocity modulus $|v_1-\bar v | = |v_2-\bar v | = v$ .
 Notice that this period increases as the solitons approach each other,
and (in this approximation) it diverges, resulting in no overlapping in momentum
space, at the classical collision time $t = x_0/v$. 

We focus on the spatial interference as the solitons move. Before
they are close enough to have a significant overlapping,
the spatial interference is approximated by
\begin{align}
I(x,t) =&|\psi|^2- |\psi_1|^2 - |\psi_2|^2 \approx\nonumber\\
 & 2|\psi_1 | |\psi_2 | \cos[\,k_{\rm dB} (x-\bar v t) - 2\nu \bar \omega t +\Theta],
\label{eq:interference}
\end{align}
where $\nu = (\mu_1-\mu_2 )/(2\hbar\bar \omega)$, and $k_{\rm dB} =2m(v_1-v_2)/\hbar$ is the
de Broglie wavenumber corresponding to the de Broglie
wave length $\lambda_{\rm dB} = 2\pi/|k_{\rm dB}| = \pi\hbar/(m v)$. The
interference manifests as an oscillatory process, $I=F\,\cos\phi$, characterized by an envelope wave 
$F (x, t) = 2|\psi_1| |\psi_2 |$ times a
carrier wave $\cos[\phi(x, t)]$, whose phase can be recast as
\begin{align}
\phi=k_{\rm dB}\bar\xi\left(\frac{x-\bar v t}{\bar \xi}-\frac{\nu}{V}\bar\omega t\right)  + \Theta,
\label{eq:relative_phase}
\end{align}
where we have introduced the non-dimensional parameter 
$V = v \,m\bar\xi/\hbar= k_{\rm dB} \bar \xi/2$,
which measures the relative velocity in intrinsic units $\hbar/(m\bar\xi)$.
From Eq. (\ref{eq:interference}--\ref{eq:relative_phase})
one can see that when the solitons have equal frequencies 
$\mu_1/\hbar = \mu_2 /\hbar$, that is $\nu = 0$, an interference
pattern arises, and it is static in the moving frame with coordinates $x'=x-\bar v t$, as given by $I = F\, \cos(k_{\rm dB} x' + \Theta)$. The pattern
is observable, roughly, if $k_{\rm dB} \bar\xi >1$; otherwise,
 for $k_{\rm dB} \bar\xi <1$ one would observe a net (single fringe)
constructive or destructive interference according to the
relative phase $\Theta$. Overall, the relative phase just shifts,
both in momentum and physical space, the positions of
interference fringes. On the other hand, if the relative
velocity vanishes, $v = 0$, one expects a time periodic pattern (as it is the case in
 bound soliton states) oscillating with a frequency $2\nu\bar\omega$.
Far from these limit cases the interference evolves into an intermediate dynamical regime
according with the ratio $|\nu/V|$ between the two dynamical
parameters $\nu$ and $V$.

In regard with the interference amplitude, keeping
the assumption of small soliton overlapping, it can
be approximated by $F (x, t) = 2 A_1 A_2 \,\mbox{sech}\, \varphi_1\, \mbox{sech}\,\varphi_2$,
where $\varphi_1=(x + x_0 -
v_1 t)/\xi_1$ and  $\varphi_2=(x - x_0 -v_2 t)/\xi_2$.
By making use of the identities between hyperbolic functions,
\begin{align}
F (x, t) = \frac{4\, A_1 A_2}{\cosh (\varphi_1+\varphi_2)+\cosh(\varphi_1-\varphi_2)},
\label{eq:interfamplitude}
\end{align}
where
\begin{align}
\varphi_1\pm\varphi_2=\frac{\xi_2\pm\xi_1}{\xi_1\,\xi_2}(x-\bar v t)+
\frac{\xi_2\mp\xi_1}{\xi_1\,\xi_2}(x_0- v t).
\label{eq:envelope_phase}
\end{align}
In the absence of current-dependent interactions $(\xi_2+\xi_1)/(\xi_1\,\xi_2)=2/\bar \xi$, $(\xi_2-\xi_1)/(\xi_1\,\xi_2)=-2\alpha/\bar\xi$, and
$A_1 A_2 =\bar A^2\,(1 - \alpha^2)$, where $\alpha= (N_1 - N_2)/(2\bar N)$ is the
differential number of particles, and $\bar A=[\bar N/(2\bar \xi)]^{1/2}$ is the average soliton amplitude.

In regular solitons, the spatial interference enters the equation of motion as a potential term $g_{\mbox{\tiny 1D}}\,I(x,t)$ whose spatial modulation, similar to a lattice potential with a a time-varying depth and spatial period determined by the linear superposition of solitons through $\cos\phi$, is expected to induce a corresponding modulation in the system state during its nonlinear evolution (see Appendix \ref{sec:currentInterf}).
In this way the interaction between solitons can be
understood as an interference process, where attractive forces reflect constructive interference, and repulsive forces reflect destructive interference. 
The analysis of the exact two-soliton solutions in regular solitons reveals that this is in fact the case \cite{Zhao2016,Zhao2017}. 

In the presence of current-density interactions, an additional interference term associated with the current-density interaction $\hbar\kappa\,J$ enters the equation of motion as a potential term.
 At the same order of approximation as Eq. (\ref{eq:interference}), it becomes 
\begin{align}
I_\kappa(x,t)\approx & \frac{\bar \xi}{\hbar}\,[\,\Re(\psi_1^*\hat p \,\psi_2)+\Re(\psi_2^*\hat p \,\psi_1)\,] = \nonumber\\
&2|\psi_1 | |\psi_2 | \left[\bar V\,\cos\phi+ W(x,t)\sin\phi\right],
\label{eq:Ikappa}
\end{align} 
where $\bar V=\bar v\,m\bar\xi/\hbar$ is the non-dimensional average velocity, and $W(x,t)$ is a space- and time-dependent amplitude that changes its sign at the collision time (see Appendix \ref{sec:currentInterf} for details). 
The interplay of the two oscillatory components in phase quadrature gives rise to a phase shift with respect to the density [see Eq.(\ref{eq:interference})] that is expected to translate into a reduction in the oscillation amplitude. In addition, the varying amplitude $W(x,t)$ introduces new time frequencies in the carrier wave.
Before elaborating in this direction, and in order to get further insight, we
revisit the collisions between regular solitons.

\begin{figure*}[htb]
	\includegraphics[width=\linewidth]{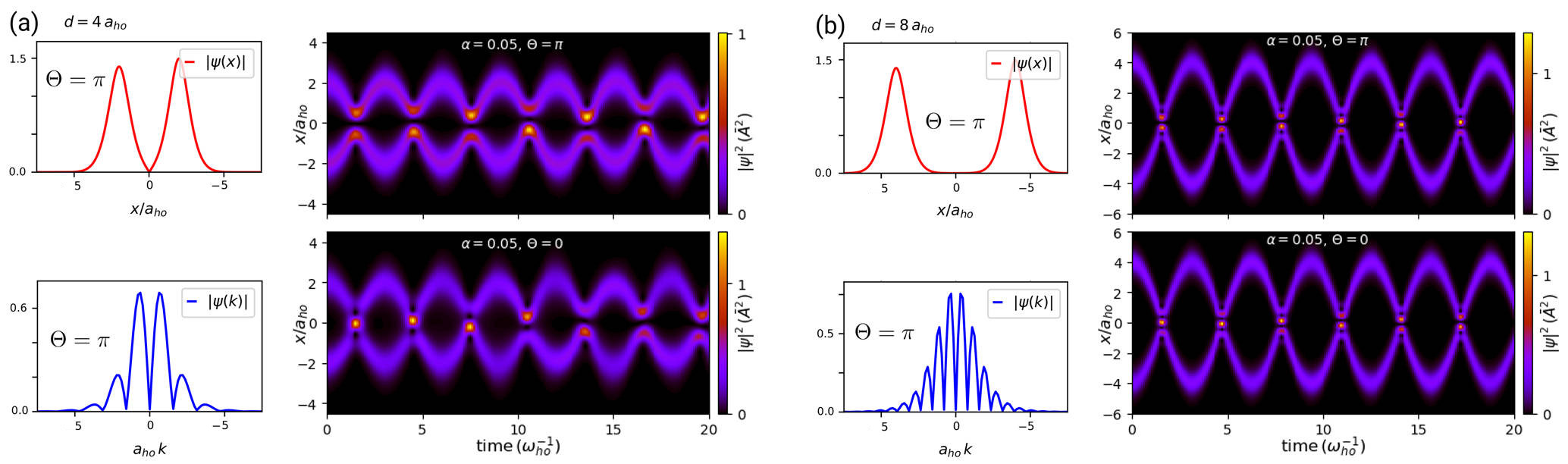}
	\caption{Collisions between regular solitons ($\kappa = 0$) in a harmonic trap.
		Panels (a) and (b) show the initial states, amplitude profiles in physical and momentum spaces, and the time evolution of the density profile for different separation and relative
		phase. The solitons are featured by the differential number of particles $\alpha = 0.05$, while the relation between the trap and the interatomic interaction is fixed by the
		length ratio $a_{\rm ho} /\xi = 1.25$.
	} 
	\label{fig:trap_coll}
\end{figure*}

\subsection{Collisions within the nonlinear Schr\"odinger equation}

In the absence of both external potential and current-dependent interaction, $U_{\rm ext}=\kappa=0$, the seminal work of Gordon \cite{Gordon1983} revealed the nature of forces acting between nearby solitons through exact solutions to the nonlinear Schr\"odinger equation.
This fact allows us to test the approximate expressions Eqs. (\ref{eq:interference}--\ref{eq:interfamplitude}). The intersoliton interaction depends on three parameters featuring the differences between solitons, namely $(\alpha, V, \Theta)$.
In this case, $\alpha$ characterizes not only the differential
number of particles, but, likewise, the relative amplitude
$ \Lambda = (A_1 -A_2 )/(2 \bar A )=\alpha$ and the relative frequency
$\nu=-\alpha$.

In many occasions, a simplified analysis of soliton collisions based on just one (usually $\Theta$) or two (usually $V$ and $\Theta$) of these parameters is presented, which, assuming solitons with equal amplitudes $\alpha=0$, leads to an oversimplified conclusion: in-phase solitons, $\Theta=0$, experience attractive forces, and opposite phase solitons, $\Theta=\pi$, experience repulsive forces between them. However, a deeper analysis shows a far richer scenario \cite{Satsuma1974,Gordon1983,Desem1987,Zhao2017}. The interaction forces decay exponentially with the soliton distance \cite{Gordon1983}, and, when the solitons are within the force reach, two dynamical regimes that depend on the ratio $|\nu/V|=\alpha/V$ can be observed.
For $\alpha/V\gg 1$ the soliton interactions involve an oscillatory dynamics characterized by two frequencies: one is directly proportional to the differential number of particles $\alpha$, and characterizes the oscillations of the soliton amplitudes, whereas the other frequency is directly proportional to $V$, and characterizes the exponential decay in the amplitude of the oscillations. 
On the other hand, for $\alpha/V\ll 1$   wave interference phenomena are dominant, and interference fringes of wave length $\lambda_{\rm dB}$ are observed.

 In both collisional regimes, the soliton interactions (the proper collisions) take place mainly during the time interval $\Delta t\approx[t_c-\tau,\,t_c+\tau]$ for a typical time $\tau = (2V\bar\omega)^{-1}$ around the collision time $t_c$ (see below); before and after this time interval the solitons translate freely, conserving the properties, amplitudes and velocities, fixed by the initial conditions. The singular case with $v=0$ gives rise to soliton bound states, whose oscillations depends on the initial inter-soliton distance \cite{Gordon1983,Desem1987,Zhao2017}.

The collision dynamics can be summarized by the time evolution of the maximum amplitude in the system, which is well approximated by the expression
\begin{align}
\max \left[\frac{|\psi(x,t)|}{\bar A}\right] = 1+\alpha+\,f(t)  \, \cos(2\alpha\,\bar\omega\,t+\Theta_0),
\label{eq:colleq1}
\end{align}
with an envelope function $f(t)$  given by
\begin{align}
f(t)= f_0
\begin{cases}  \exp{(-2V\,\bar\omega|t-t_c|)}, & \mbox{if } V < \alpha \\  \mbox{sech}[\,2V\,\bar\omega(t-t_c)], & \mbox{if } V \ge \alpha \end{cases}  
\label{eq:colleq2} \\
\mbox{and} \quad f_0= 
\begin{cases}  1-\alpha, & \mbox{if } \cos[2\alpha\,\bar\omega(t-t_c)] \ge 0  \\  2\alpha, &  \mbox{if } \cos[2\alpha\,\bar\omega(t-t_c)] < 0 , \end{cases}
\nonumber
\end{align} 
where $\Theta_0$ is a phase shift, and $t_c=(x_0+ \Delta x)/(\bar\xi \bar\omega V)$ is the collision time obtained from the initial intersoliton distance $d_0=2x_0$ and the soliton-interaction displacement $2\Delta x=-\ln[(V^2+1)/(V^2+\alpha^2)]\,\bar\xi/(1-\alpha^2)$ \cite{Gordon1983}; this latter term is not captured by the estimate Eq. (\ref{eq:interfamplitude}).
Figure \ref{fig:GPcollision} shows three examples of soliton collisions that illustrate the oscillatory regime at low relative velocity (left and middle panels), and the interference regime at high relative velocity (right panels). 
The time evolution of density and phase is depicted in the top and middle panels, whereas the bottom panels show the intersoliton distance (left), and the maximum amplitude (middle and right), comparing the numerical solution with the analytical results given by Eqs. (\ref{eq:colleq1}-\ref{eq:colleq2}). As can be seen, these equations provide a faithful characterization of the dynamics.

\subsubsection{Soliton collisions under harmonic confinement}

Apart from the influence of the trap on single soliton amplitudes, a major influence is exerted on the dynamics of soliton collisions. Assuming that the solitons are prepared in an initial state with zero velocity at symmetric positions around the trap center $x_1=-x_0$ and $x_2=x_0$, the oscillator force pushes the solitons to meet at the potential minimum, where their relative velocity, proportional to the initial separation $d_0=2x_0$, reaches the maximum value $2v=2\omega_{\rm ho}\,x_0$. 
 As before, this velocity $v$ can be compared with the intrinsic velocity $ \hbar/m\bar\zeta$, as determined from the average number of particles $\bar N$ (see Sec. \ref{sec:harmonic}) to give $V=\omega_{\rm ho} x_0\,m\bar\zeta/\hbar=  x_0\,\bar\zeta/ a_{\rm ho}^2$.
 Similarly to the untrapped case, the parameter $|\nu/V|$ determines the dynamical regime of the soliton collisions; however, an important difference arises because now $|\nu|\neq|\alpha|$, and the frequency difference $(\mu_1-\mu_2)/\hbar$, as can be inferred from Eq. (\ref{eq:mutrap}), is a nonlinear function of the differential number of particles $\alpha$.
 Still, when  $|\nu/V|>1$ the system enters an oscillatory regime, whereas for $|\nu/V|\le 1$ the collisions are featured by the presence of interference fringes. 
 Figure \ref{fig:trap_coll} shows characteristic examples of soliton collisions in a harmonic trap.
 The number of particles has been fixed by $\alpha = 0.05$,
 while both the relative phase and the intersoliton distance (hence the eventual relative velocity) are varied.
 As can be seen, the repeated collisions induced by the
 trap force do not show identical outcomes (the motion
 is quasi-periodic), due to the different soliton frequency;
 had we kept $\alpha= 0$, we would have obtained a real periodic dynamics. As anticipated, the higher the initial
 separation, as in panels (b), the clearer the interference
 pattern.

\subsection{Collisions subject to current-density interactions}

The conservation principles Eqs. (\ref{eq:momentum}--\ref{eq:energy}), along with the additional interference terms due to the current-density interaction, rule the collision dynamics.
In analogy with the amplitude interference of regular solitons, the "current interference" of chiral solitons induces an oscillatory dynamics (see Appendix \ref{sec:currentInterf} for details). Due to the velocity-dependent amplitude of the chiral solitons, the characteristic parameters of a collision change accordingly.
The differential amplitude $\Lambda$ becomes a function of $\alpha$ and the
soliton velocities $v_1 = \bar v+ v$ and $v_2 = \bar v- v$:
\begin{align}
\Lambda=\frac{1+\alpha}{2}\sqrt{1-\bar\kappa V}-\frac{1-\alpha}{2}\sqrt{1+\bar\kappa V},
\label{eq:D_amp}
\end{align}
where $\bar \kappa= \kappa\bar N/2$ provides a reference value for the current interaction strength in the system, and, as before, $V=v\,m\bar\xi/\hbar$.
The product $\bar \kappa V= \hbar \kappa v/(|g_{\mbox{\tiny 1D}}+\hbar\kappa \bar v|)=v/|v_g - \bar v|$ is a relative velocity
measure with respect to the average velocity $|v_g - \bar v|$, where $0 \le v < v_g -\bar v$ is the condition for the solitons to exist.
High relative velocities such that $v/|v_g - \bar v|\rightarrow 1$ indicate the high broadening and reduced amplitude of the forward moving soliton.
 Collisions with equal amplitudes $\Lambda = 0$,
for given $\alpha$ and $\bar v$, correspond to the relative velocity
$v_0 = |v_g - \bar v| \,2\alpha/(1 + \alpha^2)$.

\begin{figure}[tb]
	\centering
	\flushleft (a) \\
	\includegraphics[width=0.95\linewidth]{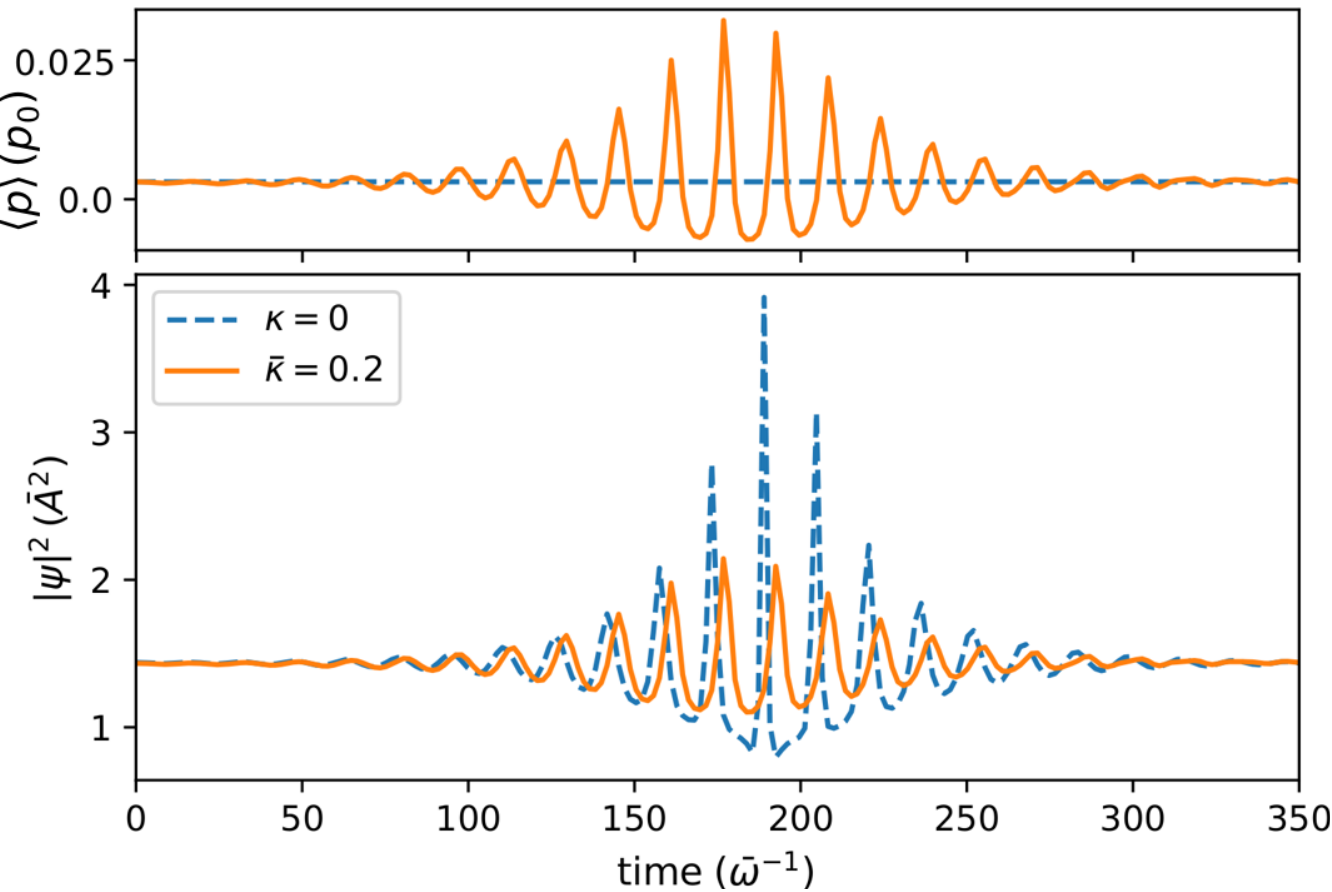}\\ \vspace{-0.5cm}
	\flushleft (b)\\
	\includegraphics[width=1.0\linewidth]{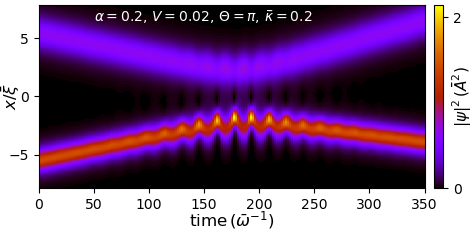}
	\caption{Chiral-soliton collisions at low current-density interactions ($\bar\kappa V=0.004$) characterized by the non-dimensional parameters $\hbar^2\bar\kappa /(m g_{\mbox{\tiny 1D}}\bar\xi)=0.2$, $\alpha=0.2$,  $V=0.02$, and global relative phase $\Theta=\pi$. The average velocity of the solitons is zero. Panel (a) shows the evolution of the system canonical momentum (top), along with the evolution of the maximum density (bottom); for comparison, the corresponding evolution for regular solitons is depicted with dashed lines. Panel (b) shows the evolution of the density profile.}
	\label{fig:jlowV}
\end{figure}

Similarly, the differential frequency can be written as a function of $\alpha$ and $\bar\kappa V$:
\begin{align}
\nu=(\bar\kappa V-\alpha)(1-\alpha \bar \kappa V).
\label{eq:nu}
\end{align}
Thus, equal soliton frequencies $\mu_1 /\hbar = \mu_2/\hbar$,
which implies also equal soliton widths $\xi_1 = \xi_2$ , are obtained when 
$\alpha = v/|v_g - \bar v|$.
 For vanishing $\bar \kappa V$, which is achieved not only for $\kappa = 0$ but also for
zero relative velocity $V = 0$ (while $\bar v$ need not to vanish),
the equalities $\Lambda = \alpha = -\nu $ of regular solitons are recovered.
The ratio $\nu/V=(\bar \kappa-\alpha/V)(1-\alpha\bar \kappa V)$, determining the oscillatory and interference dynamical regimes, involves now the three non-dimensional parameters 
 $(\alpha, \,V,\,\bar\kappa)$.

Finally, the interference envelope wave $F (x, t) = 2|\psi_1| |\psi_2 |$, as given by Eqs. (\ref{eq:interfamplitude}--\ref{eq:envelope_phase}), is obtained
with  $(\xi_2+\xi_1)/(\xi_1\,\xi_2)=2(1-\alpha\,\bar\kappa V)/\bar \xi$, $(\xi_2-\xi_1)/(\xi_1\,\xi_2)=2(\bar\kappa V-\alpha)/\bar\xi$, and
 $A_1 A_2 =\bar A^2\,(1 - \alpha^2)\sqrt{ 1 - (\bar \kappa V )^2}$.
 Therefore, the amplitude of the interference process is at least decreased in a factor 
 $(1- \bar v/v_g ) \sqrt{1 - (\bar \kappa V )^2}$ with respect to regular solitons.
  In this regard, the current-dependent interparticle interactions reduce the soliton interactions.
As we will see later, due to the phase shift between particle density and current density during
the nonlinear evolution of the system, Eqs. (\ref{eq:interference}-\ref{eq:Ikappa}), a
further soliton interaction reduction can be observed in collisions at low relative velocity.

\subsubsection{Collisions at low current-density interactions}

This dynamical regime corresponds to $\bar \kappa V \ll 1$, that is to 
$v \ll |v_g - \bar v|$. In this case, the differential amplitude Eq. (\ref{eq:D_amp})
is approximated by $\Lambda \approx \alpha - \bar \kappa V /2$, and the differential
frequency is parameterized by $\nu \approx  \bar \kappa V-\alpha$.
Since $\nu/V =   \bar \kappa-\alpha/V$, similar dynamics as in the absence of
current-density interactions is expected when $\bar \kappa\ll \alpha/V$.
Figure \ref{fig:jlowV} shows the outcome of chiral soliton collisions in this latter situation, with
low current-density interactions. We have set a zero average velocity $\bar v=0$, an interaction ratio $\hbar^2\bar\kappa /(m g_{\mbox{\tiny 1D}}\bar\xi)=0.2$, a differential number of particles $\alpha=0.2$, and a relative velocity determined by $V=0.02$. 
As expected the results are qualitatively similar to those in regular solitons [shown for comparison in panel (a) by dashed lines], and the corresponding dynamical regime characterized by amplitude oscillations can be observed. 
As predicted, the amplitude oscillation period is practically indistinguishable from the
case of regular solitons. Despite the canonical momentum is not a conserved quantity, small variations are observed, and the system recovers the total initial canonical momentum after the collision event.

Though, relevant differences with respect to regular solitons appear, as the peak density reduction (of about 50 $\%$ in this case), and the emergence of
 partial or even total reflection during the collisions. It is worth comparing Fig \ref{fig:jlowV}(b), which shows a total reflection of chiral solitons, with the central panels of Fig. \ref{fig:GPcollision} for regular solitons. The latter solitons go through each other in a collision, and no reflection is produced. On the contrary, chiral solitons collisions involve in general both transmission and reflection processes that are regulated by the relative phase. 
 Our results show that total reflection occurs at low relative velocity $|\nu/V|\gg 1$, whereas total transmission can be observed at high relative velocity $|\nu/V|\ll 1$.
 In this latter regime, as far as low current-density interactions $\bar\kappa\ll1$ are kept, chiral-soliton collisions are even more similar to those of regular solitons, with no significant amplitude reduction, and the appearance of interference fringes of wave number $k_{\rm dB}=2m\,v/\hbar$ (see Appendix \ref{sec:currentInterf} for details). 

\begin{figure}[htb]
	\centering
	\flushleft (a) \\
	\includegraphics[width=0.9\linewidth]{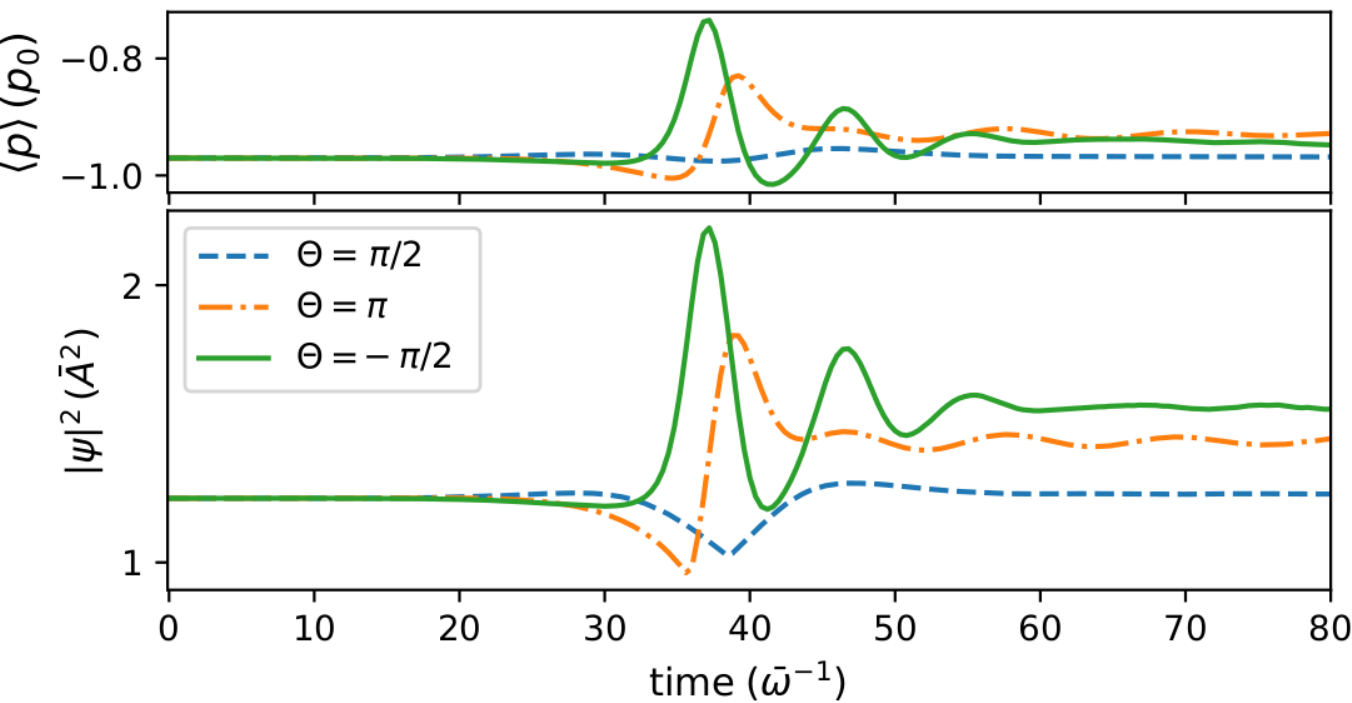}\\ \vspace{-0.5cm}
	\flushleft (b)\\
	\includegraphics[width=1.0\linewidth]{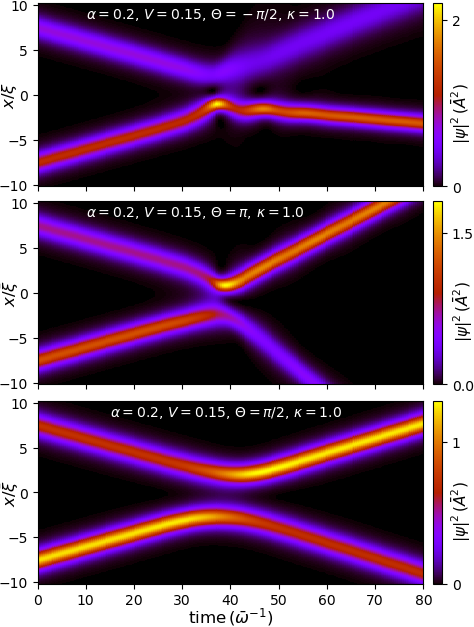}
	\caption{Chiral-soliton collisions with no contact interactions ($g_{\mbox{\tiny 1D}}$=0), intermediate current-density interaction $\bar\kappa V=0.15$, and varying initial relative phases $\Theta$, as viewed in a reference frame moving with the average soliton velocity $\bar V=\bar v\,m\bar\xi/\hbar=1$.	As in Fig. \ref{fig:jlowV}, 
		panels (a) show the evolution of the system canonical momentum and maximum density, whereas panel (b) shows the evolution of the density profile.}
	\label{fig:jmidV}
\end{figure}
\begin{figure}[htb]
	\centering
	\flushleft (a) \\
	\includegraphics[width=0.9\linewidth]{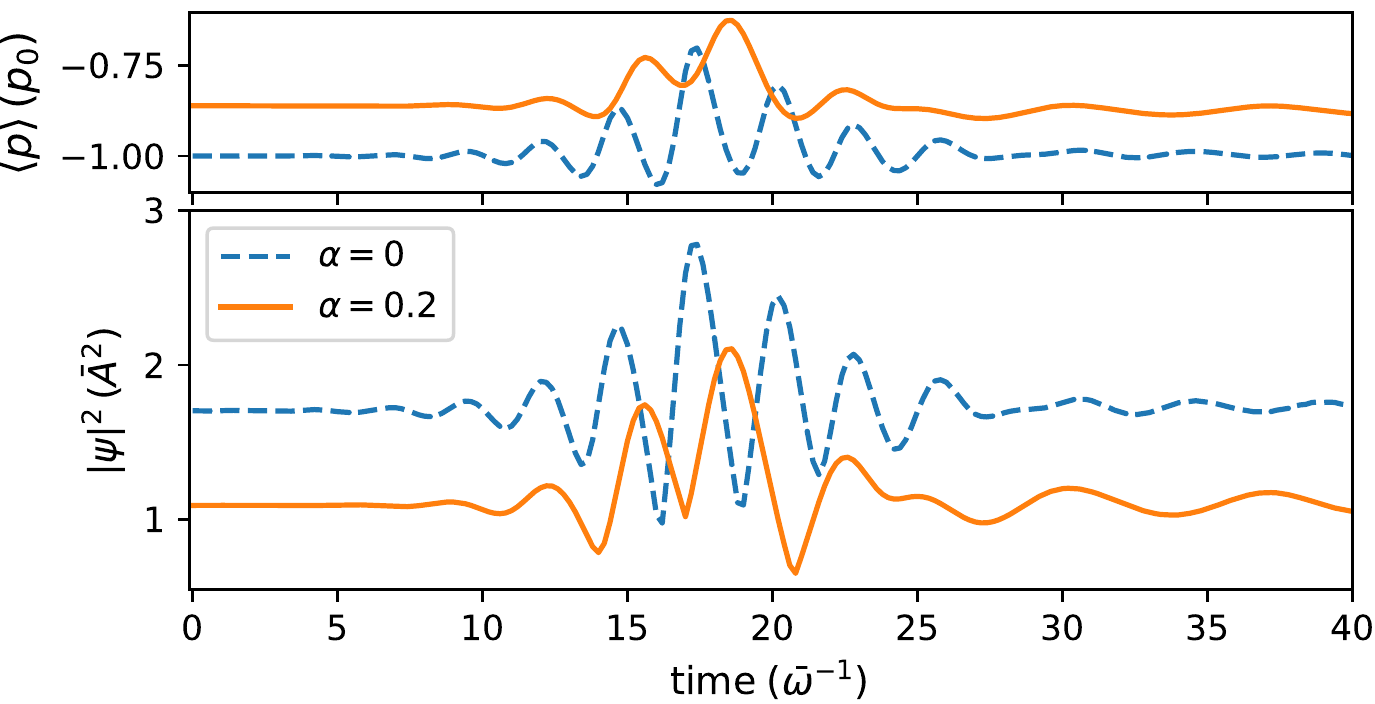}\\ \vspace{-0.5cm}
	\flushleft (b)\\
	\includegraphics[width=1.0\linewidth]{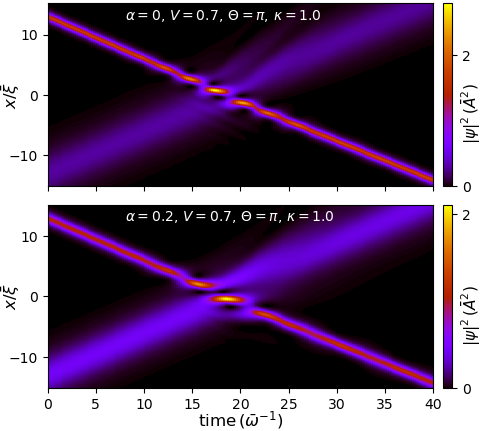}
	\caption{Same as Fig. \ref{fig:jmidV} but for high current-density interaction $\bar\kappa V=0.7$ and varying differential number of particles $\alpha$. The differential amplitude and frequency are $\Lambda=-0.38$ and $\nu=0.7$ at $\alpha=0$, versus  $\Lambda=-0.19$ and  $\nu=0.43$ at $\alpha=0.2$. }
	\label{fig:jhighV}
\end{figure}

\subsubsection{General collisions }

To explore chiral soliton collisions at higher current-density interaction,
we have chosen a reference frame moving with the average velocity $\bar v$, so
that the solitons have equal relative-velocity modulus
$|v_1 - \bar v| = |v_2 - \bar v| = v$. Viewed as a scattering event
of two incoming solitons, the collision produces outgoing waves that can be classified under two main sets: one set is characterized by two outgoing solitons with, in general, different velocities and amplitudes 
from the incoming ones; the second set includes outgoing waves that involve non-solitonic radiation along with solitons.
 In both sets, the initial relative phase has a strong influence in
the scattering process (significantly stronger than in regular
solitons), so that, for otherwise equal initial
soliton parameters, different relative phases can change
the outcome of the collision from one set to the other.
This picture is consistent with the results of Ref. \cite{Dingwall2018},
where collisions between solitons with equal number of
particles, $\alpha = 0$, were addressed, and the elasticity of the
collision versus the relative phase was measured through
the restitution coefficient (as a ratio of incoming and outgoing kinetic energies).

Figure \ref{fig:jmidV} shows chiral soliton collisions for an intermediate value of the current-density interaction $\bar\kappa V=v/\bar v=0.15$, no contact interaction ($g_{\mbox{\tiny 1D}}=0$), and varying relative phase $\Theta$. The average soliton velocity is $\bar V=\bar v\,m\bar\xi/\hbar=1$, and the differential number of particles is $\alpha=0.2$, so that the differential amplitude and frequency become $\Lambda=0.12$ and $ \nu=-0.05$, respectively. 
Although the ratio $|\nu/V|=1/3$ points to a non-oscillatory dynamics, 
the sizable influence of the relative phase gives rise to different scenarios.
While at $\Theta=-\pi/2$ one can see the almost total reflection of the solitons (with a 2$\%$ variation in each soliton particle number), at $\Theta=\pi/2$ the almost total transmission (with practically conserved canonical momentum) is observed. In between, a highly asymmetrical outcome is produced at $\Theta=\pi$ (with an outgoing differential number of particles $\alpha=0.4$), which, due to the conservation of the total mechanical momentum $\Pi$, Eq. (\ref{eq:momentum}), involves a significant change in the velocities of the outgoing solitons.
The peak density achieved during the collisions is equally affected by the relative phase, with very small variation for the total transmission event. 
Particular values of the relative phase close to the transition from total reflection to total transmission can extend the duration of the collision through oscillation cycles mediated by momentum and particle exchange (see Appendix \ref{sec:currentInterf}).

High current-density interaction, as shown in Fig. \ref{fig:jhighV} for $\bar\kappa V=0.7$ and average soliton velocity $\bar V=\bar v\,m\bar\xi/\hbar=1$, leads to the almost full transmission of the solitons through the collision, along with the appearance of interference fringes. The relative phase becomes less relevant, since it only changes the position of the maxima and minima of the interference fringes.  Though, the particular arrangement of the fringes is involved in the amount of non-solitonic radiation that can also be observed in this regime. The presence of radiation can be understood as related to the
generation of nonlinear waves that exceed the limit speed $|\bar v|$ (in general $|v_g-\bar v|$) during the scattering event.

\subsubsection{Current-density interaction and harmonic trapping}

   As in regular solitons, we focus on initial states with two solitons situated symmetrically around the trap center. The initial soliton separation determines the soliton speeds at collision time. Additionally, the harmonic force induces repeated collisions at twice the harmonic frequency. The oscillator length scale and the average number of particles per soliton are chosen to give $\bar\xi=0.8\, a_{\rm ho}$. Since the initial state is made of static solitons (located at the turning points of the subsequent evolution), both contact and current-dependent interactions are switched on. The strength of the latter is characterized by the velocity $v_g=g_{\mbox{\tiny 1D}}/(\hbar\kappa)$.
 
\begin{figure}[htb]
	\includegraphics[width=1.0\linewidth]{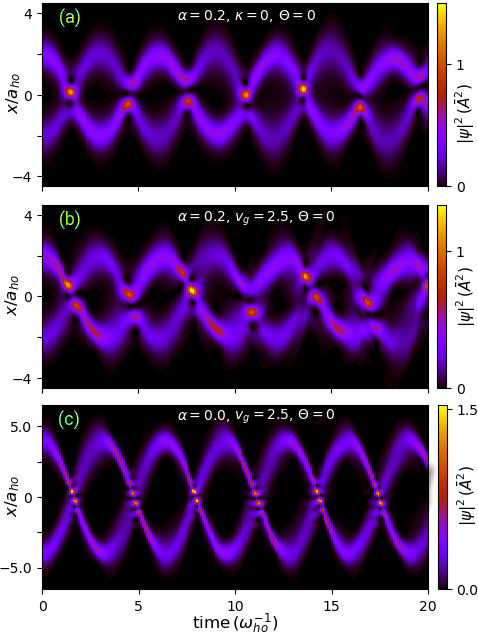}
	\caption{Chiral-solitons collisions in a harmonic trap. Panel (a), for regular solitons ($\kappa=0$), is shown for comparison. In panels (b)-(c),
		the current-density interaction is characterized by the velocity $v_g=g_{\mbox{\tiny 1D}}/(\hbar\kappa)=2.5\,\hbar/(m\bar\xi)$.  } 
	\label{fig:jtrap}
\end{figure}

 Figure \ref{fig:jtrap} shows several examples for varying parameters. For comparison, panel (a) depicts a case with $\kappa=0$, differential number of particles $\alpha=0.2$, and short soliton separation $d=4\,a_{\rm ho}$ (that produces a small overlapping of their tails); as a consequence, the repeated collisions show different outcomes.
 Once again, the relative phase does not produce a qualitatively different dynamics. 
 Panel (b) illustrates the system time evolution for these same parameters plus current-density interaction parameterized by $v_g=2.5\,\hbar/(m\bar\xi)$. A more complex scenario arises due to the inelastic character of the collisions, and the dynamics become more irregular at longer times.
 By increasing the soliton separation, as in panel (c), where $d=8\,a_{\rm ho}$ the maximum velocity at the trap center $\omega_{\rm ho}d/(2v_g)=1.28$ brings the forward moving solitons into
 a temporarily unstable state (for $x\in[-0.875,\,0.875]\,a_{\rm ho}$ their effective interparticle interaction is repulsive). Nevertheless the oscillator force is capable to balance the dispersive effects, and the system shows repeated cycles with the characteristic interference patterns of high velocity collisions. 

\section{Conclusions}
\label{sec:Conclusions}

We have presented a general analysis of chiral soliton collisions in Bose-Einstein condensates subject to current-dependent interaction. By varying the differential amplitude, the relative phase, the average velocity, and the relative velocity of the two solitons, the soliton collision dynamics have been discussed extensively. We characterize the different dynamical regimes that give rise to oscillatory and interference phenomena. Guided by the linear superposition of the solitons, we have determined the relevant time and space scales that characterize the observed oscillatory and interference phenomena of the collisions. The amplitude reduction with respect to the case of regular solitons has been revealed as a special feature in the chiral dynamics. Furthermore, in order to compare with feasible ultracold gas experiments, we have investigated the influence of harmonic confinement on the emergence and the interaction of chiral solitons.

\begin{acknowledgments}
This work was supported by the National Natural Science Foundation
of China (Grant No. 11402199), the Natural Science Foundation of Shaanxi Province(Grant No. 2022JM-004, No. 2018JM1050), and the Education Department Foundation of Shaanxi
Province(Grant No. 14JK1676).
\end{acknowledgments}

\appendix
\section{Units}
\label{sec:units}

 In the absence of external potential, the regular 1D GP equation written in  non-dimensional form is
\begin{align}
\left\lbrace i \frac{\partial }{\partial \tilde t}+\beta\,\frac{1}{2}\frac{\partial^2 }{\partial \tilde x^2}- \tilde g_{\mbox{\tiny 1D}} |\tilde \psi|^{2}\,\right\rbrace  \psi(x,t),
\label{eq:Ggpe}
\end{align}
where $\tilde x=x/ \ell$ and time $\tilde t=\omega t$ are the dimensionless coordinates, $\tilde \psi=\sqrt{\ell/\gamma}\,\psi$, with $\gamma$ constant, is the non-dimensional wave function, and the non-dimensional parameters $\beta$ and $\tilde g_{\mbox{\tiny 1D}}$ depend on the selection of units of length $\ell$ and time $\omega^{-1}$:
\begin{align}
\beta=\frac{\hbar^2}{m\ell^2\,\hbar\omega},\quad \tilde g_{\mbox{\tiny 1D}}= \frac{g_{\mbox{\tiny 1D}}}{\hbar\omega\ell}\gamma.
\label{eq:units}
\end{align}
By choosing $\beta=|\tilde g_{\mbox{\tiny 1D}}|=1$, the non-dimensional GP Eq. (\ref{eq:Ggpe}) takes a universal form fixed by the units $\hbar\omega=m(|g_{\mbox{\tiny 1D}}|\gamma)^2/\hbar^2$, and $\ell=\hbar^2/(m|g_{\mbox{\tiny 1D}}|\gamma)$; the normalization becomes 
\begin{align}
\int d\tilde x |\tilde \psi(\tilde x,\tilde t)|^2=\frac{N}{\gamma},
\label{eq:norm}
\end{align}
 and the velocity is measured in units of $\omega\ell=|g_{\mbox{\tiny 1D}}|\gamma/\hbar$. For the analysis of the two-soliton system we have chosen $\gamma=N/4=\bar N/2$, where $\bar N$ is the average number of particles per soliton. With this choice $\int d\tilde x |\tilde \psi|^2=4$, and the unit of length matches the width of the average soliton $\ell=\bar \xi=2\hbar^2/(m|g_{\mbox{\tiny 1D}}|\bar N)$, that is a soliton containing the average number of particles; the corresponding energy unit is $\hbar\omega=\hbar^2/m\bar\xi^2=2\bar\mu\equiv\hbar\bar\omega$, where $\bar \mu/\hbar$ is the characteristic frequency of the average soliton.  

Analogously, if there is only current-density interaction, the equation of motion 
written in  non-dimensional form is
\begin{align}
\left\lbrace i \frac{\partial }{\partial \tilde t}+\beta\,\frac{1}{2}\frac{\partial^2 }{\partial \tilde x^2}-  \tilde J\,\right\rbrace  \psi(x,t),
\label{eq:Jgpe}
\end{align}
where  $\tilde J=\kappa J/\omega$ is the non-dimensional current density
 Since $\kappa$ is dimensionless, only one parameter, $\beta$, determines the units, which, with $\beta=1$, fulfill $\hbar\omega=\hbar^2/(m\ell^2)$. With no extra parameter introducing a fixed scale unit, Eq. (\ref{eq:Jgpe}) is scale invariant \cite{Jackiw1997}. 
However, the two-soliton system introduces two velocity scales, the average, $\bar v$, and the relative, $2v$, soliton velocities, with the constraint $v/|\bar v|< 1$. In this case, with the same normalization factor as before, $\gamma=\bar N/2$, we choose the units  
 such that $\omega\ell=\hbar/(m\ell)=\kappa \gamma \bar v$, and then $\hbar\omega=m\,(\kappa \gamma \bar v)^2$.

 In the presence of both current-density $\kappa\neq0$ and contact $g_{\mbox{\tiny 1D}}\neq 0$ interaction, we rewrite the non-dimensional Eq. (\ref{eq:Ggpe}) with $\tilde g=|g_{\mbox{\tiny 1D}}+\hbar\kappa \bar v|\gamma/(\hbar\omega\ell)$, so that, by setting $\tilde g=\beta=1$ and $\gamma= \bar N/2$ for the two-soliton system, the resulting units are $\hbar\omega=m(|g_{\mbox{\tiny 1D}}+\hbar\kappa \bar v| \gamma/\hbar)^2$, and $\ell=2\hbar^2/(m|g_{\mbox{\tiny 1D}}+\hbar\kappa \bar v| \bar N)$, in analogy with the case with only contact interactions. Thus, the velocity unit is $\omega\ell=|g_{\mbox{\tiny 1D}}+\hbar\kappa \bar v| \bar N/(2\hbar)=\kappa\gamma|v_g-\bar v|$, where $v_g=|g_{\mbox{\tiny 1D}}|/(\hbar\kappa)$. Notice that $v_g>\bar v$ is a necessary condition for the soliton existence.

\section{ Collisions of two chiral solitons} 
\label{sec:currentInterf}

\subsection{Amplitude and current-density mediated interference} 

\begin{figure}[tb]
	\centering
	\flushleft (a) \\
	\includegraphics[width=1.0\linewidth]{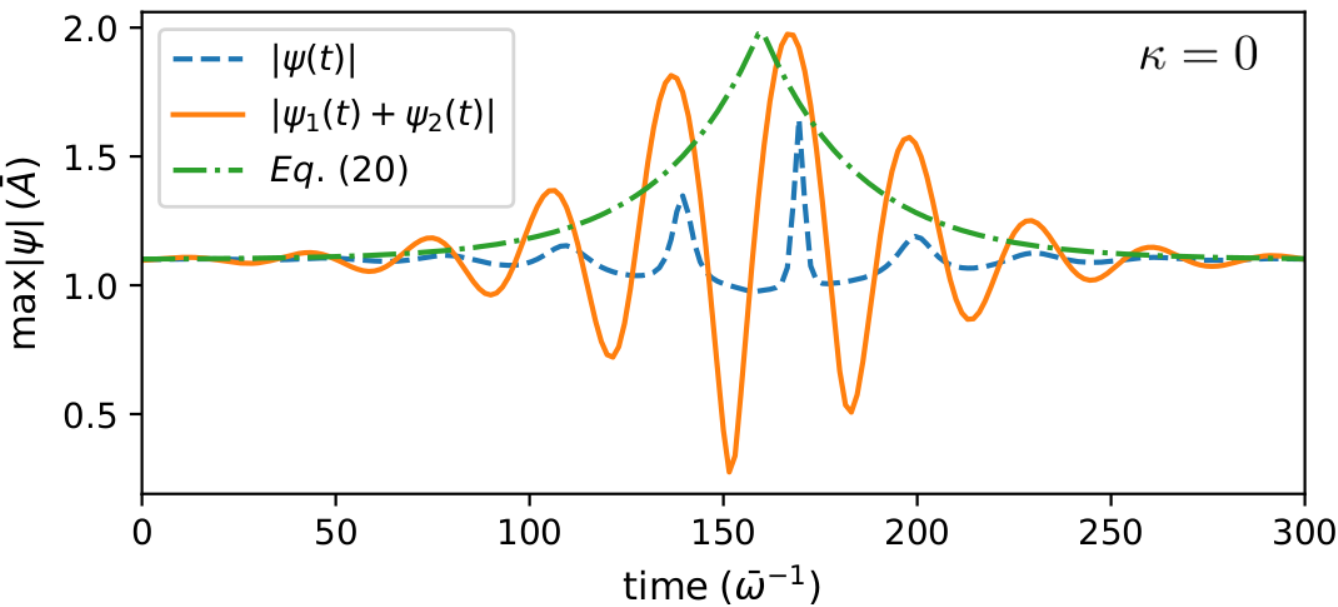}\\ \vspace{-0.5cm}
	\flushleft (b)\\
	\includegraphics[width=1.0\linewidth]{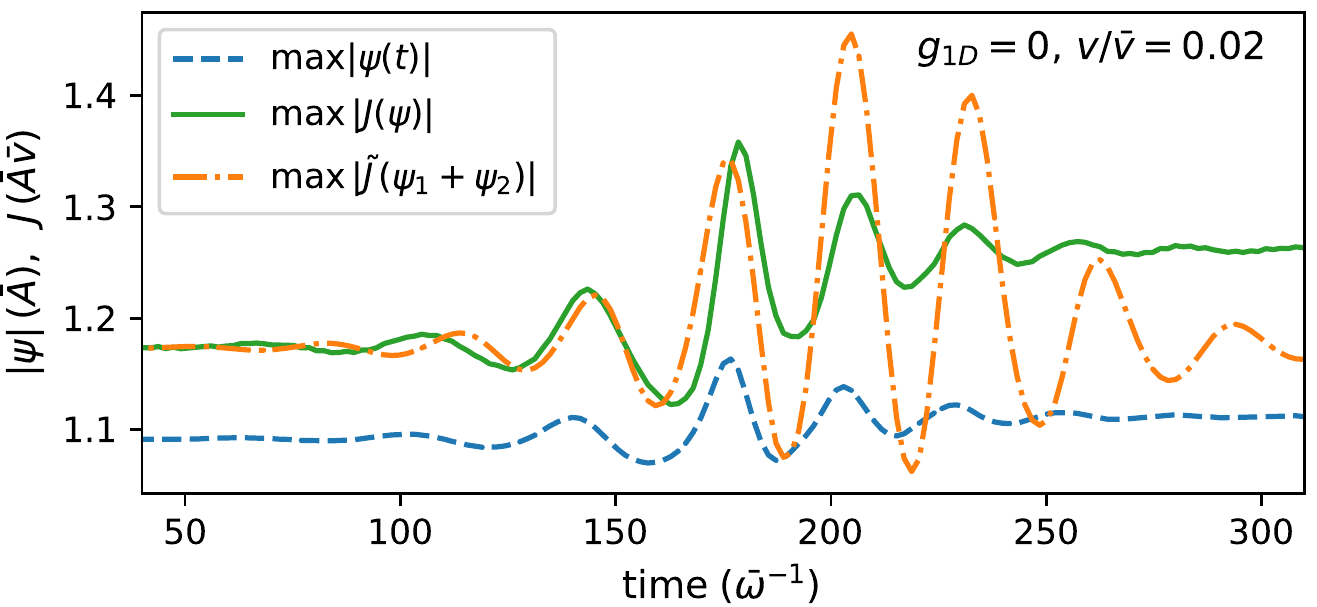}\\ \vspace{-0.5cm}
	\flushleft (c)\\
	\includegraphics[width=1.0\linewidth]{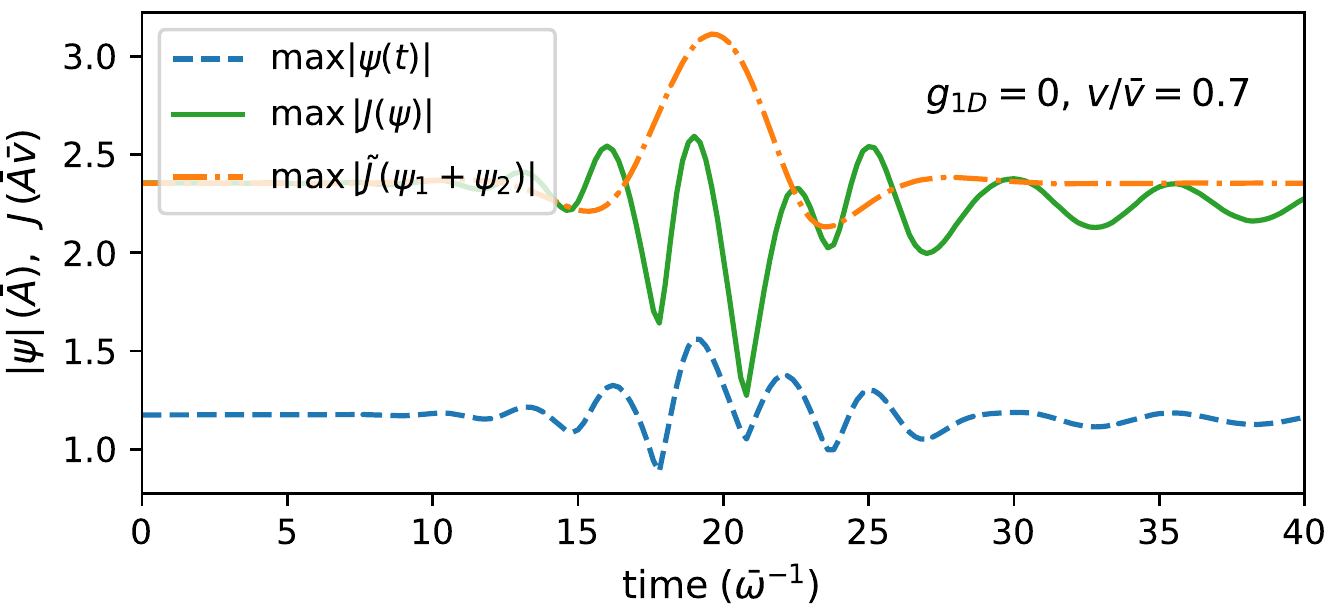}
	\caption{(a) Nonlinear evolution of the maximum amplitude (dashed line) in a two-regular-soliton system characterized by  $\alpha=0.1,\,V=0.02$, and $m\bar\xi\,g_{\mbox{\tiny 1D}}/\hbar^2=1$. The evolution of non-interacting solitons (solid line) is shown for comparison.
		(b) Nonlinear evolution of the current density (solid line) and maximum amplitude (dashed line) of chiral solitons  with $m\bar\xi\,\kappa\,\bar v/\hbar=1$, no contact interaction ($g_{\mbox{\tiny 1D} }=0$), and otherwise same parameters as panel (a). The evolution of the current density of non-interacting solitons (dot-dashed line) is shown for comparison. 
	(c) Same as panel (b) but for higher relative velocity. }
	\label{fig:linear}
\end{figure}

The superposition of the soliton wave functions led to the amplitude interference Eq. (\ref{eq:interference}), which is expected to remain as a good approximation during the nonlinear time evolution of the solitons while they show no significant overlapping. The soliton interference enters the dynamics through the mean-field, contact-interaction term in GP Eq. (\ref{eq:gpe}).
Interestingly, in regular solitons the time evolution of the non-interacting,  interfering solitons, as driven by Eq. (\ref{eq:interference}), captures the characteristic time and length scales of the nonlinear time evolution.
 Figure \ref{fig:linear}(a) shows an example of regular soliton collisions at low relative velocity, $\alpha/V=5$, where these features are compared by means of the evolution of the system maximum amplitude. Although the amplitude predicted by the non-interacting solitons (solid line) is manifestly higher, the period of the amplitude oscillations, and the duration of the collision (roughly, the time during which interference is significant) are the same.

Analogously, the current-density interaction $\hbar\kappa J=\hbar\kappa\Re(\psi^*\hat p \,\psi)/m$ introduces additional interference terms in the two-chiral-soliton dynamics, associated with the coupling of amplitude and momentum of different solitons, that we write as
$\hbar^2\kappa/(m\bar\xi)\,I_\kappa(x,t)$, where
 $I_\kappa=(\bar\xi/\hbar)[\Re(\psi^*\hat p \,\psi)-\Re(\psi_1^*\hat p \,\psi_1)-\Re(\psi_2^*\hat p \,\psi_2)]$. When there is no significant soliton overlapping, at the same order of approximation as Eq. (\ref{eq:interference}), it becomes
\begin{align}
I_\kappa\approx &\, (\bar\xi/\hbar)\,[\,\Re(\psi_1^*\hat p \,\psi_2)+\Re(\psi_2^*\hat p \,\psi_1)\,] =
\nonumber \\
& 2|\psi_1 | |\psi_2 | \left[\bar V\,\cos\phi+ W(x,t)\sin\phi\right].
\label{eq:jInterf}
\end{align}
Here $\phi=\,k_{\rm dB} (x-\bar v t) - 2\nu \bar \omega t +\Theta$ is the relative phase, as defined in Eq. (\ref{eq:relative_phase}), $\bar V=\bar v\,m\bar\xi/\hbar$ is the non-dimensional average velocity, and 
\begin{align}
W(x,t)=\bar\xi\left(\frac{\tanh\,\varphi_2}{2\xi_2}-\frac{\tanh\,\varphi_1}{2\xi_1}\right),
\label{eq:jInterf_W}
\end{align}
where $\varphi_2=(x-x_0-v_2 t)/\xi_2$ and $\varphi_1=(x+x_0-v_1 t)/\xi_1$
The cosine part of Eq. (\ref{eq:jInterf}), proportional to the average velocity $\bar V$, is in-phase with the amplitude interference of expression
Eq. (\ref{eq:interference}), whereas the sine part provides a quadrature term with 
space and time varying amplitude $W$.
 For well resolved solitons, the latter quantity can be approximated before the collision, by $W_0\approx -(\xi_1+\xi_2)\,\bar\xi/(2\xi_1\,\xi_2)=-(1-\alpha \bar\kappa V)$ for the space between solitons, and $W_0\approx (\xi_1-\xi_2)\,\bar\xi/(2\xi_1\,\xi_2)=(\alpha-\bar\kappa V)$ otherwise; after the collision time, $W$ reverses its sign, so it experiences an overall change of $2|W_0|$  during an interval of the order of $\bar \xi/v$ around the collision time.

Equation (\ref{eq:jInterf}) can be recast as
\begin{align}
I_\kappa\approx \,  2|\psi_1 | |\psi_2 | \sqrt{\bar V^2+ W^2}\,\cos\left(\phi-\arctan\frac{W}{\bar V}\right),
\label{eq:jInterf1}
\end{align}
where it is explicitly stated that the potential introduced by the current-density interaction is dephased by the time-varying phase  $\arctan (W/\bar V)$, and scaled by the time-varying amount $ (\bar V^2+ W^2)^{1/2}$,   with respect to the particle density, Eq. (\ref{eq:interference}).

Figure \ref{fig:linear}(b) shows an example of chiral soliton collisions in the absence of contact interactions, and otherwise equal parameters as Fig. \ref{fig:linear}(a). The average velocity is set by $m\bar\xi\,\kappa\,\bar v/\hbar=1$, so that the differential amplitude and frequency are $\Lambda=0.09$ and $\nu=-0.08$, respectively. The latter quantity produces a ratio $|\nu/V|=4$ that brings the system in the oscillatory regime. The current of non-interacting solitons (represented with a reduced and phase shifted amplitude, $\tilde J$, by a dot-dashed line) provides a good approximation to the time frequency of the real dynamics. However, as can be seen in Fig. \ref{fig:linear}(c), at high relative velocity $v/\bar v=0.7$ the linear approximation fails to provide a characteristic frequency of the collision due to a pulsating dynamics accompanied by
non-solitonic radiation. The differential amplitude and frequency are $\Lambda=-0.28$ and $\nu=0.56$, opposite to the corresponding values at low relative velocity.

\subsubsection{Interference fringes at low interaction}

\begin{figure}[tb]
	\includegraphics[width=1.0\linewidth]{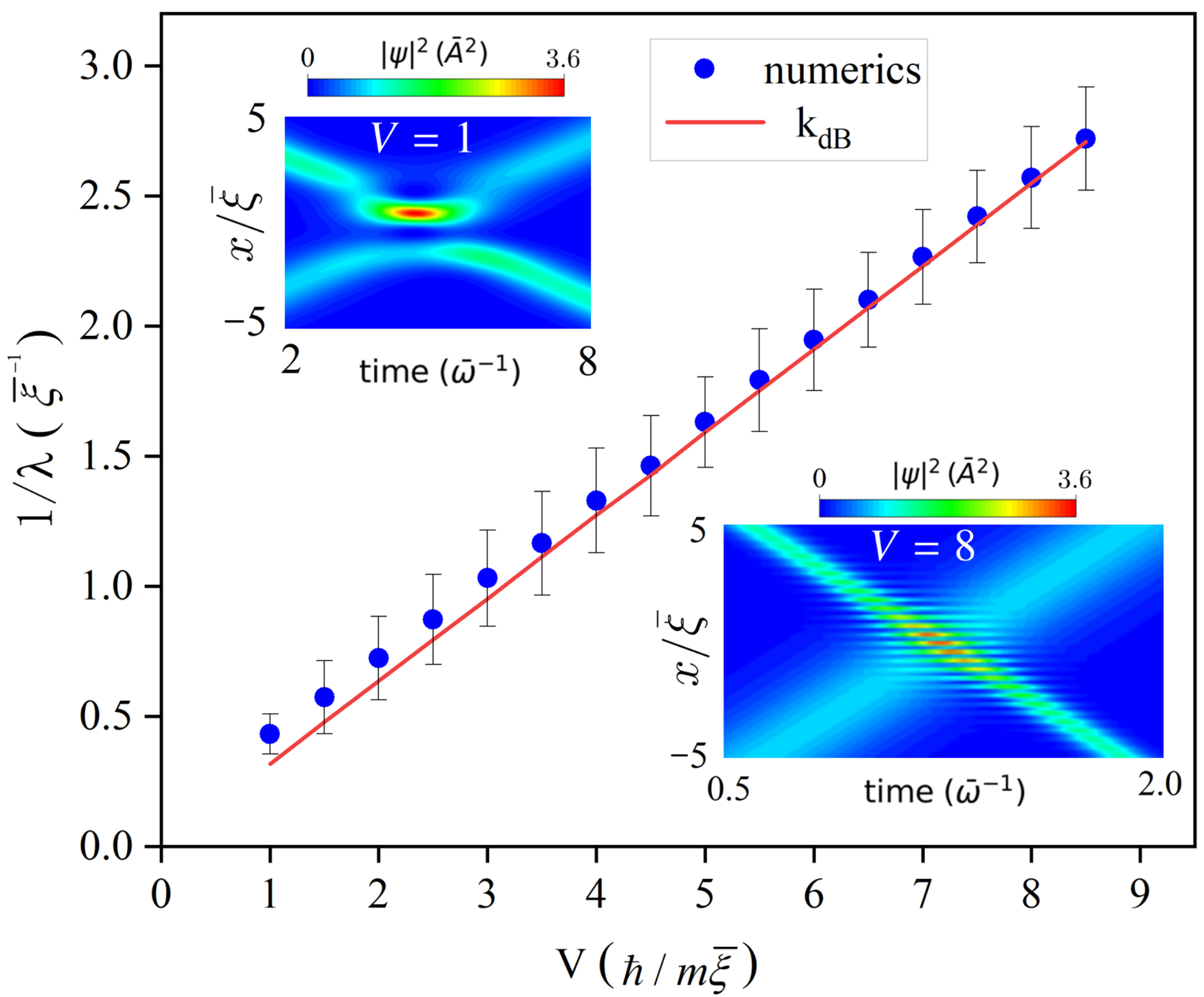}
	\caption{Distance $\lambda$ between interference fringes in counter propagating chiral soliton collisions as a function of the relative velocity parameterized by $V=m\bar\xi\,v/\hbar$. Both contact and current-density interactions are present, with the ratio $\hbar^2\kappa\bar N/(m\bar\xi\,g_{\mbox{\tiny 1D}})=0.1$. The insets show the time evolution of the density profiles for two velocity values $V=1,\,8$.} 
	\label{fig:trans2}
\end{figure}

As in regular solitons, the interference fringes that emerge from chiral soliton collisions at low current-density interactions $\bar\kappa V \ll 1$ are determined by the de-Broglie wavenumber associated with the relative velocity $k_{\rm dB}=2mv/\hbar$. Figure \ref{fig:trans2} shows our numerical results in this regime for chiral soliton collisions in the presence of both contact and current-density interactions. 
The determination of the distance between fringes has been obtained from a Fourier analysis of the data, so that the filled circles correspond to the wavenumber with (non-zero, local) maximum amplitude, and the error bars indicate the width of the local maximum.

\subsubsection{Influence of the relative phase}

\begin{figure}[tb]
		\centering
	\flushleft (a) \\
	\includegraphics[width=0.9\linewidth]{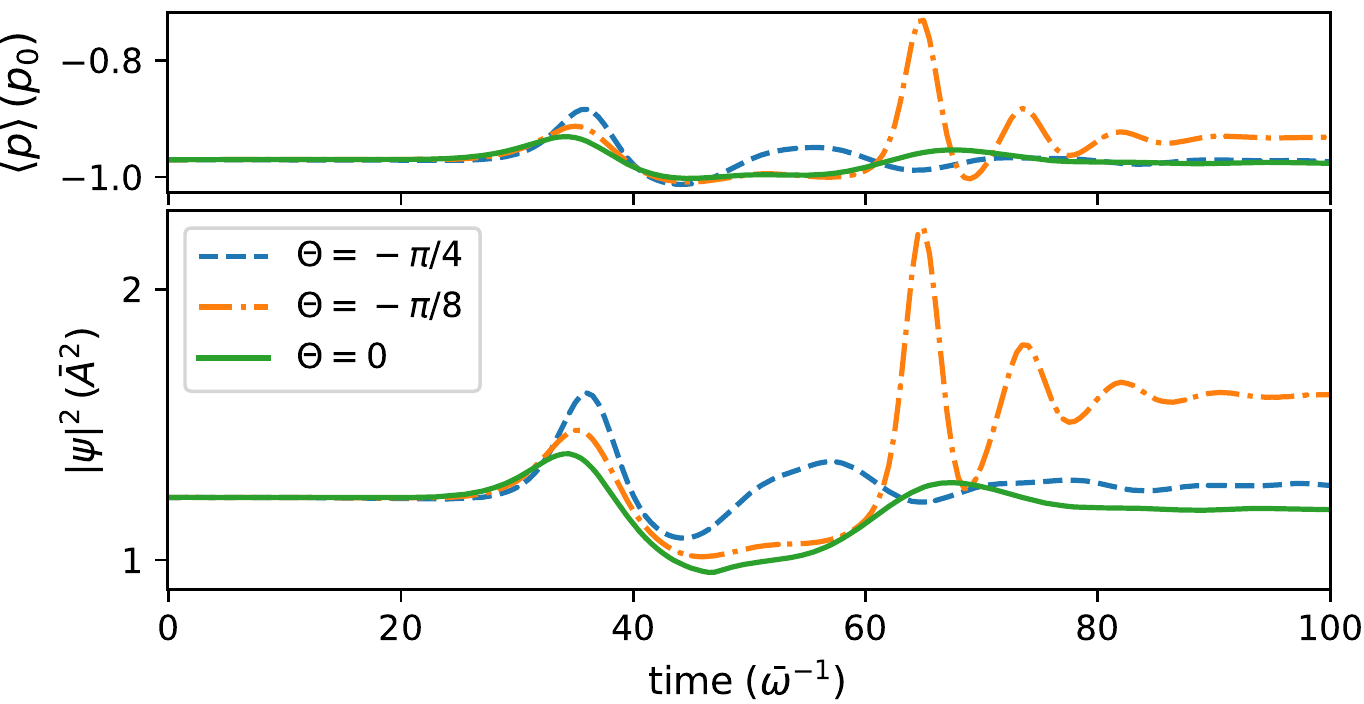}\\ \vspace{-0.5cm}
	\flushleft (b)\\
	\includegraphics[width=1.0\linewidth]{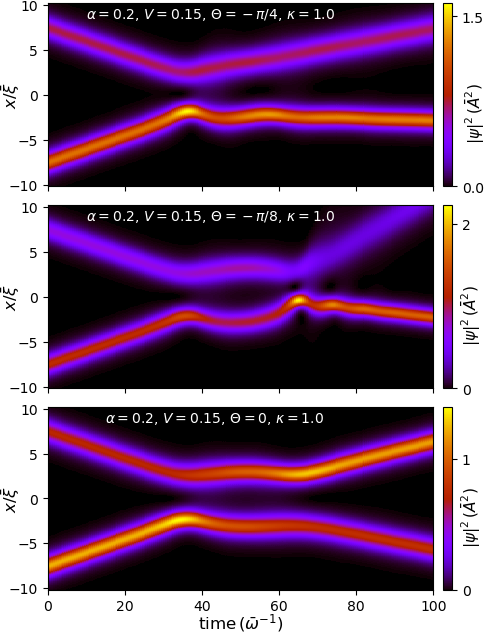}
	\caption{Same as Fig. \ref{fig:jmidV} for different values of the initial, global relative phase $\Theta$ close to the transition from predominant reflection (top panel) to predominant transmission (bottom panel) events. } 
	\label{fig:trans1}
\end{figure}

The relevance of the relative phase in chiral soliton collisions
 can be clearly seen at intermediate values of the current-density interaction, as shown in Fig. \ref{fig:jmidV} in the main text. 
  Particular values of the relative phase
close to the transition from total reflection to total transmission cause a significant variation in the duration of the collision. As can be seen in Fig. \ref{fig:trans1}, one or more cycles of soliton oscillations, mediated by momentum and particle exchange, can be observed.

\subsection{Energy and momentum} 

The lack of conservation of the canonical momentum allows for the exchange of
momentum and interaction energy between solitons; from Eq. (\ref{eq:momentum})
\begin{align}
P_t-P_0=\frac{\hbar\kappa}{2}\int dx\,(|\psi_t|^4-|\psi_0|^4),
\label{eq:PiPf}
\end{align}
where $P_0$ and $P_t$  are the total canonical momentum
of the initial $\psi_0$ and final $\psi_t$ states, respectively.
For the initial solitons one obtains a canonical momentum
$P_0 = m(N_1 v_1 + N_2v_2 ) = \bar N\hbar(2\bar V+ \alpha V )/\bar \xi$,
and a momentum contribution from the gauge field $\bar \kappa \int dx |\psi_0 |^4 /\bar N= 2\hbar \kappa \bar N(1-\alpha^2)(1+\alpha\bar\kappa V )/(3\bar \xi)$,
which for the simplest case of equal number of particles
$\alpha = 0$ gives $P_0 = 2\hbar \bar V \bar N /\bar \xi$ and $2\hbar\kappa\bar N/(3\bar\xi)$,
respectively.

The energy conservation Eq. (\ref{eq:energy}) states that $E_t =
E_0$, where the initial two-soliton energy is $E_0 =\sum_j 
N_j [ mv_j^2 /2 -\mu_j /3- |g_{\mbox{\tiny 1D}}| A_j^2/3 ]$, for $j = 1, 2$, and $\mu_j=-\hbar^2/(2m\xi_j^2)$. 
As a function of the collision parameters, the conserved energy is
\begin{align}
E_t=\bar N\hbar \bar\omega\left\lbrace 
\bar V^2+V^2+\alpha\bar V V+\frac{2}{3}(1+\alpha\nu) \right. \nonumber\\ \left.
-\frac{2}{3}\left\lvert\frac{g_{\mbox{\tiny 1D}}}{g_{\bar v}}\right\rvert (1+\Lambda^2+2\alpha\Lambda)
\right\rbrace .
\label{eq:EiEt}
\end{align}

From Eq. (\ref{eq:PiPf}), one can see that the change in canonical momentum is accompanied by a
change in the density distribution (or alternatively, in the number of particles) of the solitons. This variation is however limited by the conservation of energy Eq. (\ref{eq:EiEt}), which in the absence of contact interaction is just the conservation of the total kinetic energy.

\bibliography{sol_col_dens_dep_gauge}

\begin{thebibliography}{34}%
\makeatletter
\providecommand \@ifxundefined [1]{%
 \@ifx{#1\undefined}
}%
\providecommand \@ifnum [1]{%
 \ifnum #1\expandafter \@firstoftwo
 \else \expandafter \@secondoftwo
 \fi
}%
\providecommand \@ifx [1]{%
 \ifx #1\expandafter \@firstoftwo
 \else \expandafter \@secondoftwo
 \fi
}%
\providecommand \natexlab [1]{#1}%
\providecommand \enquote  [1]{``#1''}%
\providecommand \bibnamefont  [1]{#1}%
\providecommand \bibfnamefont [1]{#1}%
\providecommand \citenamefont [1]{#1}%
\providecommand \href@noop [0]{\@secondoftwo}%
\providecommand \href [0]{\begingroup \@sanitize@url \@href}%
\providecommand \@href[1]{\@@startlink{#1}\@@href}%
\providecommand \@@href[1]{\endgroup#1\@@endlink}%
\providecommand \@sanitize@url [0]{\catcode `\\12\catcode `\$12\catcode
  `\&12\catcode `\#12\catcode `\^12\catcode `\_12\catcode `\%12\relax}%
\providecommand \@@startlink[1]{}%
\providecommand \@@endlink[0]{}%
\providecommand \url  [0]{\begingroup\@sanitize@url \@url }%
\providecommand \@url [1]{\endgroup\@href {#1}{\urlprefix }}%
\providecommand \urlprefix  [0]{URL }%
\providecommand \Eprint [0]{\href }%
\providecommand \doibase [0]{http://dx.doi.org/}%
\providecommand \selectlanguage [0]{\@gobble}%
\providecommand \bibinfo  [0]{\@secondoftwo}%
\providecommand \bibfield  [0]{\@secondoftwo}%
\providecommand \translation [1]{[#1]}%
\providecommand \BibitemOpen [0]{}%
\providecommand \bibitemStop [0]{}%
\providecommand \bibitemNoStop [0]{.\EOS\space}%
\providecommand \EOS [0]{\spacefactor3000\relax}%
\providecommand \BibitemShut  [1]{\csname bibitem#1\endcsname}%
\let\auto@bib@innerbib\@empty
\bibitem [{\citenamefont {Sackett}\ \emph {et~al.}(1999)\citenamefont
  {Sackett}, \citenamefont {Gerton}, \citenamefont {Welling},\ and\
  \citenamefont {Hulet}}]{Sackett1999}%
  \BibitemOpen
  \bibfield  {author} {\bibinfo {author} {\bibfnamefont {C.}~\bibnamefont
  {Sackett}}, \bibinfo {author} {\bibfnamefont {J.~M.}\ \bibnamefont {Gerton}},
  \bibinfo {author} {\bibfnamefont {M.}~\bibnamefont {Welling}}, \ and\
  \bibinfo {author} {\bibfnamefont {R.~G.}\ \bibnamefont {Hulet}},\ }\href
  {\doibase 10.1103/PhysRevLett.82.876} {\bibfield  {journal} {\bibinfo
  {journal} {Phys. Rev. Lett.}\ }\textbf {\bibinfo {volume} {82}},\ \bibinfo
  {pages} {876} (\bibinfo {year} {1999})}\BibitemShut {NoStop}%
\bibitem [{\citenamefont {Gerton}\ \emph {et~al.}(2000)\citenamefont {Gerton},
  \citenamefont {Strekalov}, \citenamefont {Prodan},\ and\ \citenamefont
  {Hulet}}]{Gerton2000}%
  \BibitemOpen
  \bibfield  {author} {\bibinfo {author} {\bibfnamefont {J.~M.}\ \bibnamefont
  {Gerton}}, \bibinfo {author} {\bibfnamefont {D.}~\bibnamefont {Strekalov}},
  \bibinfo {author} {\bibfnamefont {I.}~\bibnamefont {Prodan}}, \ and\ \bibinfo
  {author} {\bibfnamefont {R.~G.}\ \bibnamefont {Hulet}},\ }\href {\doibase
  10.1038/35047030} {\bibfield  {journal} {\bibinfo  {journal} {Nature}\
  }\textbf {\bibinfo {volume} {408}},\ \bibinfo {pages} {692} (\bibinfo {year}
  {2000})}\BibitemShut {NoStop}%
\bibitem [{\citenamefont {Donley}\ \emph {et~al.}(2001)\citenamefont {Donley},
  \citenamefont {Claussen}, \citenamefont {Cornish}, \citenamefont {Roberts},
  \citenamefont {Cornell},\ and\ \citenamefont {Wieman}}]{Donley2001}%
  \BibitemOpen
  \bibfield  {author} {\bibinfo {author} {\bibfnamefont {E.~A.}\ \bibnamefont
  {Donley}}, \bibinfo {author} {\bibfnamefont {N.~R.}\ \bibnamefont
  {Claussen}}, \bibinfo {author} {\bibfnamefont {S.~L.}\ \bibnamefont
  {Cornish}}, \bibinfo {author} {\bibfnamefont {J.~L.}\ \bibnamefont
  {Roberts}}, \bibinfo {author} {\bibfnamefont {E.~A.}\ \bibnamefont
  {Cornell}}, \ and\ \bibinfo {author} {\bibfnamefont {C.~E.}\ \bibnamefont
  {Wieman}},\ }\href {https://www.nature.com/articles/35085500} {\bibfield
  {journal} {\bibinfo  {journal} {Nature}\ }\textbf {\bibinfo {volume} {412}},\
  \bibinfo {pages} {295} (\bibinfo {year} {2001})}\BibitemShut {NoStop}%
\bibitem [{\citenamefont {Strecker}\ \emph {et~al.}(2002)\citenamefont
  {Strecker}, \citenamefont {Partridge}, \citenamefont {Truscott},\ and\
  \citenamefont {Hulet}}]{Strecker2002}%
  \BibitemOpen
  \bibfield  {author} {\bibinfo {author} {\bibfnamefont {K.~E.}\ \bibnamefont
  {Strecker}}, \bibinfo {author} {\bibfnamefont {G.~B.}\ \bibnamefont
  {Partridge}}, \bibinfo {author} {\bibfnamefont {A.~G.}\ \bibnamefont
  {Truscott}}, \ and\ \bibinfo {author} {\bibfnamefont {R.~G.}\ \bibnamefont
  {Hulet}},\ }\href {\doibase 10.1038/nature747} {\bibfield  {journal}
  {\bibinfo  {journal} {Nature}\ }\textbf {\bibinfo {volume} {417}},\ \bibinfo
  {pages} {150} (\bibinfo {year} {2002})}\BibitemShut {NoStop}%
\bibitem [{\citenamefont {{Al Khawaja}}\ \emph {et~al.}(2002)\citenamefont {{Al
  Khawaja}}, \citenamefont {Stoof}, \citenamefont {Hulet}, \citenamefont
  {Strecker},\ and\ \citenamefont {Partridge}}]{AlKhawaja2002}%
  \BibitemOpen
  \bibfield  {author} {\bibinfo {author} {\bibfnamefont {U.}~\bibnamefont {{Al
  Khawaja}}}, \bibinfo {author} {\bibfnamefont {H.}~\bibnamefont {Stoof}},
  \bibinfo {author} {\bibfnamefont {R.~G.}\ \bibnamefont {Hulet}}, \bibinfo
  {author} {\bibfnamefont {K.}~\bibnamefont {Strecker}}, \ and\ \bibinfo
  {author} {\bibfnamefont {G.}~\bibnamefont {Partridge}},\ }\href {\doibase
  10.1103/PhysRevLett.89.200404} {\bibfield  {journal} {\bibinfo  {journal}
  {Phys. Rev. Lett.}\ }\textbf {\bibinfo {volume} {89}},\ \bibinfo {pages}
  {200404} (\bibinfo {year} {2002})}\BibitemShut {NoStop}%
\bibitem [{\citenamefont {Cornish}\ \emph {et~al.}(2006)\citenamefont
  {Cornish}, \citenamefont {Thompson},\ and\ \citenamefont
  {Wieman}}]{Cornish2006}%
  \BibitemOpen
  \bibfield  {author} {\bibinfo {author} {\bibfnamefont {S.~L.}\ \bibnamefont
  {Cornish}}, \bibinfo {author} {\bibfnamefont {S.~T.}\ \bibnamefont
  {Thompson}}, \ and\ \bibinfo {author} {\bibfnamefont {C.~E.}\ \bibnamefont
  {Wieman}},\ }\href {\doibase 10.1103/PhysRevLett.96.170401} {\bibfield
  {journal} {\bibinfo  {journal} {Phys. Rev. Lett.}\ }\textbf {\bibinfo
  {volume} {96}},\ \bibinfo {pages} {170401} (\bibinfo {year}
  {2006})}\BibitemShut {NoStop}%
\bibitem [{\citenamefont {Nguyen}\ \emph {et~al.}(2014)\citenamefont {Nguyen},
  \citenamefont {Dyke}, \citenamefont {Luo}, \citenamefont {Malomed},\ and\
  \citenamefont {Hulet}}]{Nguyen2014}%
  \BibitemOpen
  \bibfield  {author} {\bibinfo {author} {\bibfnamefont {J.~H.~V.}\
  \bibnamefont {Nguyen}}, \bibinfo {author} {\bibfnamefont {P.}~\bibnamefont
  {Dyke}}, \bibinfo {author} {\bibfnamefont {D.}~\bibnamefont {Luo}}, \bibinfo
  {author} {\bibfnamefont {B.~A.}\ \bibnamefont {Malomed}}, \ and\ \bibinfo
  {author} {\bibfnamefont {R.~G.}\ \bibnamefont {Hulet}},\ }\href {\doibase
  10.1038/nphys3135} {\bibfield  {journal} {\bibinfo  {journal} {Nat. Phys.}\
  }\textbf {\bibinfo {volume} {10}},\ \bibinfo {pages} {918} (\bibinfo {year}
  {2014})}\BibitemShut {NoStop}%
\bibitem [{\citenamefont {Gordon}(1983)}]{Gordon1983}%
  \BibitemOpen
  \bibfield  {author} {\bibinfo {author} {\bibfnamefont {J.~P.}\ \bibnamefont
  {Gordon}},\ }\href {\doibase 10.1364/OL.8.000596} {\bibfield  {journal}
  {\bibinfo  {journal} {Opt. Lett.}\ }\textbf {\bibinfo {volume} {8}},\
  \bibinfo {pages} {596} (\bibinfo {year} {1983})}\BibitemShut {NoStop}%
\bibitem [{\citenamefont {Desem}\ and\ \citenamefont {Chu}(1987)}]{Desem1987}%
  \BibitemOpen
  \bibfield  {author} {\bibinfo {author} {\bibfnamefont {C.}~\bibnamefont
  {Desem}}\ and\ \bibinfo {author} {\bibfnamefont {P.}~\bibnamefont {Chu}},\
  }\href@noop {} {\bibfield  {journal} {\bibinfo  {journal} {IEE Proc.
  J-Optoelectron.}\ }\textbf {\bibinfo {volume} {134}},\ \bibinfo {pages} {145}
  (\bibinfo {year} {1987})}\BibitemShut {NoStop}%
\bibitem [{\citenamefont {Salasnich}\ \emph {et~al.}(2003)\citenamefont
  {Salasnich}, \citenamefont {Parola},\ and\ \citenamefont
  {Reatto}}]{Salasnich2003}%
  \BibitemOpen
  \bibfield  {author} {\bibinfo {author} {\bibfnamefont {L.}~\bibnamefont
  {Salasnich}}, \bibinfo {author} {\bibfnamefont {A.}~\bibnamefont {Parola}}, \
  and\ \bibinfo {author} {\bibfnamefont {L.}~\bibnamefont {Reatto}},\ }\href
  {\doibase 10.1103/PhysRevLett.91.080405} {\bibfield  {journal} {\bibinfo
  {journal} {Phys. Rev. Lett.}\ }\textbf {\bibinfo {volume} {91}},\ \bibinfo
  {pages} {080405} (\bibinfo {year} {2003})}\BibitemShut {NoStop}%
\bibitem [{\citenamefont {Leung}\ \emph {et~al.}(2002)\citenamefont {Leung},
  \citenamefont {Truscott},\ and\ \citenamefont {Baldwin}}]{Leung2003}%
  \BibitemOpen
  \bibfield  {author} {\bibinfo {author} {\bibfnamefont {V.~Y.~F.}\
  \bibnamefont {Leung}}, \bibinfo {author} {\bibfnamefont {A.~G.}\ \bibnamefont
  {Truscott}}, \ and\ \bibinfo {author} {\bibfnamefont {K.~G.~H.}\ \bibnamefont
  {Baldwin}},\ }\href@noop {} {\bibfield  {journal} {\bibinfo  {journal} {Phys.
  Rev. A}\ }\textbf {\bibinfo {volume} {66}},\ \bibinfo {pages} {061602}
  (\bibinfo {year} {2002})}\BibitemShut {NoStop}%
\bibitem [{\citenamefont {Carr}\ and\ \citenamefont
  {Brand}(2004{\natexlab{a}})}]{Carr2004}%
  \BibitemOpen
  \bibfield  {author} {\bibinfo {author} {\bibfnamefont {L.~D.}\ \bibnamefont
  {Carr}}\ and\ \bibinfo {author} {\bibfnamefont {J.}~\bibnamefont {Brand}},\
  }\href {\doibase 10.1103/PhysRevLett.92.040401} {\bibfield  {journal}
  {\bibinfo  {journal} {Phys. Rev. Lett.}\ }\textbf {\bibinfo {volume} {92}},\
  \bibinfo {pages} {040401} (\bibinfo {year} {2004}{\natexlab{a}})}\BibitemShut
  {NoStop}%
\bibitem [{\citenamefont {Carr}\ and\ \citenamefont
  {Brand}(2004{\natexlab{b}})}]{Carr2004a}%
  \BibitemOpen
  \bibfield  {author} {\bibinfo {author} {\bibfnamefont {L.~D.}\ \bibnamefont
  {Carr}}\ and\ \bibinfo {author} {\bibfnamefont {J.}~\bibnamefont {Brand}},\
  }\href {\doibase 10.1103/PhysRevA.70.033607} {\bibfield  {journal} {\bibinfo
  {journal} {Phys. Rev. A}\ }\textbf {\bibinfo {volume} {70}},\ \bibinfo
  {pages} {33607} (\bibinfo {year} {2004}{\natexlab{b}})}\BibitemShut {NoStop}%
\bibitem [{\citenamefont {Dabrowska-W{\"{u}}ster}\ \emph
  {et~al.}(2009)\citenamefont {Dabrowska-W{\"{u}}ster}, \citenamefont
  {W{\"{u}}ster},\ and\ \citenamefont {Davis}}]{Wuster2009}%
  \BibitemOpen
  \bibfield  {author} {\bibinfo {author} {\bibfnamefont {B.~J.}\ \bibnamefont
  {Dabrowska-W{\"{u}}ster}}, \bibinfo {author} {\bibfnamefont {S.}~\bibnamefont
  {W{\"{u}}ster}}, \ and\ \bibinfo {author} {\bibfnamefont {M.~J.}\
  \bibnamefont {Davis}},\ }\href {\doibase 10.1088/1367-2630/11/5/053017}
  {\bibfield  {journal} {\bibinfo  {journal} {New J. Phys.}\ }\textbf {\bibinfo
  {volume} {11}},\ \bibinfo {pages} {053017} (\bibinfo {year}
  {2009})}\BibitemShut {NoStop}%
\bibitem [{\citenamefont {Billam}\ \emph {et~al.}(2012)\citenamefont {Billam},
  \citenamefont {Marchant}, \citenamefont {Cornish}, \citenamefont {Gardiner},\
  and\ \citenamefont {Parker}}]{Billam2012}%
  \BibitemOpen
  \bibfield  {author} {\bibinfo {author} {\bibfnamefont {T.}~\bibnamefont
  {Billam}}, \bibinfo {author} {\bibfnamefont {A.}~\bibnamefont {Marchant}},
  \bibinfo {author} {\bibfnamefont {S.}~\bibnamefont {Cornish}}, \bibinfo
  {author} {\bibfnamefont {S.}~\bibnamefont {Gardiner}}, \ and\ \bibinfo
  {author} {\bibfnamefont {N.}~\bibnamefont {Parker}},\ }in\ \href@noop {}
  {\emph {\bibinfo {booktitle} {Spontaneous Symmetry Breaking, Self-Trapping,
  and Josephson Oscillations}}}\ (\bibinfo  {publisher} {Springer},\ \bibinfo
  {year} {2012})\ pp.\ \bibinfo {pages} {403--455}\BibitemShut {NoStop}%
\bibitem [{\citenamefont {Zhao}\ \emph {et~al.}(2016)\citenamefont {Zhao},
  \citenamefont {Ling}, \citenamefont {Yang},\ and\ \citenamefont
  {Liu}}]{Zhao2016}%
  \BibitemOpen
  \bibfield  {author} {\bibinfo {author} {\bibfnamefont {L.-C.}\ \bibnamefont
  {Zhao}}, \bibinfo {author} {\bibfnamefont {L.}~\bibnamefont {Ling}}, \bibinfo
  {author} {\bibfnamefont {Z.-Y.}\ \bibnamefont {Yang}}, \ and\ \bibinfo
  {author} {\bibfnamefont {J.}~\bibnamefont {Liu}},\ }\href@noop {} {\bibfield
  {journal} {\bibinfo  {journal} {Nonlinear Dyn.}\ }\textbf {\bibinfo {volume}
  {83}},\ \bibinfo {pages} {659} (\bibinfo {year} {2016})}\BibitemShut
  {NoStop}%
\bibitem [{\citenamefont {Zhao}\ \emph {et~al.}(2017)\citenamefont {Zhao},
  \citenamefont {Ling}, \citenamefont {Yang},\ and\ \citenamefont
  {Yang}}]{Zhao2017}%
  \BibitemOpen
  \bibfield  {author} {\bibinfo {author} {\bibfnamefont {L.-C.}\ \bibnamefont
  {Zhao}}, \bibinfo {author} {\bibfnamefont {L.}~\bibnamefont {Ling}}, \bibinfo
  {author} {\bibfnamefont {Z.-Y.}\ \bibnamefont {Yang}}, \ and\ \bibinfo
  {author} {\bibfnamefont {W.-L.}\ \bibnamefont {Yang}},\ }\href@noop {}
  {\bibfield  {journal} {\bibinfo  {journal} {Nonlinear Dyn.}\ }\textbf
  {\bibinfo {volume} {88}},\ \bibinfo {pages} {2957} (\bibinfo {year}
  {2017})}\BibitemShut {NoStop}%
\bibitem [{\citenamefont {Everitt}\ \emph {et~al.}(2017)\citenamefont
  {Everitt}, \citenamefont {Sooriyabandara}, \citenamefont {Guasoni},
  \citenamefont {Wigley}, \citenamefont {Wei}, \citenamefont {McDonald},
  \citenamefont {Hardman}, \citenamefont {Manju}, \citenamefont {Close},
  \citenamefont {Kuhn}, \citenamefont {Szigeti}, \citenamefont {Kivshar},\ and\
  \citenamefont {Robins}}]{Everitt2017}%
  \BibitemOpen
  \bibfield  {author} {\bibinfo {author} {\bibfnamefont {P.~J.}\ \bibnamefont
  {Everitt}}, \bibinfo {author} {\bibfnamefont {M.~A.}\ \bibnamefont
  {Sooriyabandara}}, \bibinfo {author} {\bibfnamefont {M.}~\bibnamefont
  {Guasoni}}, \bibinfo {author} {\bibfnamefont {P.~B.}\ \bibnamefont {Wigley}},
  \bibinfo {author} {\bibfnamefont {C.~H.}\ \bibnamefont {Wei}}, \bibinfo
  {author} {\bibfnamefont {G.~D.}\ \bibnamefont {McDonald}}, \bibinfo {author}
  {\bibfnamefont {K.~S.}\ \bibnamefont {Hardman}}, \bibinfo {author}
  {\bibfnamefont {P.}~\bibnamefont {Manju}}, \bibinfo {author} {\bibfnamefont
  {J.~D.}\ \bibnamefont {Close}}, \bibinfo {author} {\bibfnamefont {C.~C.~N.}\
  \bibnamefont {Kuhn}}, \bibinfo {author} {\bibfnamefont {S.~S.}\ \bibnamefont
  {Szigeti}}, \bibinfo {author} {\bibfnamefont {Y.~S.}\ \bibnamefont
  {Kivshar}}, \ and\ \bibinfo {author} {\bibfnamefont {N.~P.}\ \bibnamefont
  {Robins}},\ }\href {\doibase 10.1103/PhysRevA.96.041601} {\bibfield
  {journal} {\bibinfo  {journal} {Phys. Rev. A}\ }\textbf {\bibinfo {volume}
  {96}},\ \bibinfo {pages} {041601} (\bibinfo {year} {2017})}\BibitemShut
  {NoStop}%
\bibitem [{\citenamefont {Nguyen}\ \emph {et~al.}(2017)\citenamefont {Nguyen},
  \citenamefont {Luo},\ and\ \citenamefont {Hulet}}]{Nguyen2017}%
  \BibitemOpen
  \bibfield  {author} {\bibinfo {author} {\bibfnamefont {J.~H.~V.}\
  \bibnamefont {Nguyen}}, \bibinfo {author} {\bibfnamefont {D.}~\bibnamefont
  {Luo}}, \ and\ \bibinfo {author} {\bibfnamefont {R.~G.}\ \bibnamefont
  {Hulet}},\ }\href {\doibase 10.1126/science.aal3220} {\bibfield  {journal}
  {\bibinfo  {journal} {Science}\ }\textbf {\bibinfo {volume} {356}},\ \bibinfo
  {pages} {422} (\bibinfo {year} {2017})}\BibitemShut {NoStop}%
\bibitem [{\citenamefont {Fr\"olian}\ \emph {et~al.}(2022)\citenamefont
  {Fr\"olian}, \citenamefont {Chisholm}, \citenamefont {Neri}, \citenamefont
  {Cabrera}, \citenamefont {Ramos}, \citenamefont {Celi},\ and\ \citenamefont
  {Tarruell}}]{Frolian2022}%
  \BibitemOpen
  \bibfield  {author} {\bibinfo {author} {\bibfnamefont {A.}~\bibnamefont
  {Fr\"olian}}, \bibinfo {author} {\bibfnamefont {C.~S.}\ \bibnamefont
  {Chisholm}}, \bibinfo {author} {\bibfnamefont {E.}~\bibnamefont {Neri}},
  \bibinfo {author} {\bibfnamefont {C.~R.}\ \bibnamefont {Cabrera}}, \bibinfo
  {author} {\bibfnamefont {R.}~\bibnamefont {Ramos}}, \bibinfo {author}
  {\bibfnamefont {A.}~\bibnamefont {Celi}}, \ and\ \bibinfo {author}
  {\bibfnamefont {L.}~\bibnamefont {Tarruell}},\ }\href
  {https://www.nature.com/articles/s41586-022-04943-3} {\bibfield  {journal}
  {\bibinfo  {journal} {Nature}\ }\textbf {\bibinfo {volume} {608}},\ \bibinfo
  {pages} {293} (\bibinfo {year} {2022})}\BibitemShut {NoStop}%
\bibitem [{\citenamefont {Aglietti}\ \emph {et~al.}(1996)\citenamefont
  {Aglietti}, \citenamefont {Griguolo}, \citenamefont {Jackiw}, \citenamefont
  {Pi},\ and\ \citenamefont {Seminara}}]{Aglietti1996}%
  \BibitemOpen
  \bibfield  {author} {\bibinfo {author} {\bibfnamefont {U.}~\bibnamefont
  {Aglietti}}, \bibinfo {author} {\bibfnamefont {L.}~\bibnamefont {Griguolo}},
  \bibinfo {author} {\bibfnamefont {R.}~\bibnamefont {Jackiw}}, \bibinfo
  {author} {\bibfnamefont {S.-Y.}\ \bibnamefont {Pi}}, \ and\ \bibinfo {author}
  {\bibfnamefont {D.}~\bibnamefont {Seminara}},\ }\href {\doibase
  10.1103/PhysRevLett.77.4406} {\bibfield  {journal} {\bibinfo  {journal}
  {Phys. Rev. Lett.}\ }\textbf {\bibinfo {volume} {77}},\ \bibinfo {pages}
  {4406} (\bibinfo {year} {1996})}\BibitemShut {NoStop}%
\bibitem [{\citenamefont {Clark}\ \emph {et~al.}(2018)\citenamefont {Clark},
  \citenamefont {Anderson}, \citenamefont {Feng}, \citenamefont {Gaj},
  \citenamefont {Levin},\ and\ \citenamefont {Chin}}]{Clark2018}%
  \BibitemOpen
  \bibfield  {author} {\bibinfo {author} {\bibfnamefont {L.~W.}\ \bibnamefont
  {Clark}}, \bibinfo {author} {\bibfnamefont {B.~M.}\ \bibnamefont {Anderson}},
  \bibinfo {author} {\bibfnamefont {L.}~\bibnamefont {Feng}}, \bibinfo {author}
  {\bibfnamefont {A.}~\bibnamefont {Gaj}}, \bibinfo {author} {\bibfnamefont
  {K.}~\bibnamefont {Levin}}, \ and\ \bibinfo {author} {\bibfnamefont
  {C.}~\bibnamefont {Chin}},\ }\href {\doibase 10.1103/PhysRevLett.121.030402}
  {\bibfield  {journal} {\bibinfo  {journal} {Phys. Rev. Lett.}\ }\textbf
  {\bibinfo {volume} {121}},\ \bibinfo {pages} {030402} (\bibinfo {year}
  {2018})}\BibitemShut {NoStop}%
\bibitem [{\citenamefont {G{\"o}rg}\ \emph {et~al.}(2019)\citenamefont
  {G{\"o}rg}, \citenamefont {Sandholzer}, \citenamefont {Minguzzi},
  \citenamefont {Desbuquois}, \citenamefont {Messer},\ and\ \citenamefont
  {Esslinger}}]{Gorg2019}%
  \BibitemOpen
  \bibfield  {author} {\bibinfo {author} {\bibfnamefont {F.}~\bibnamefont
  {G{\"o}rg}}, \bibinfo {author} {\bibfnamefont {K.}~\bibnamefont
  {Sandholzer}}, \bibinfo {author} {\bibfnamefont {J.}~\bibnamefont
  {Minguzzi}}, \bibinfo {author} {\bibfnamefont {R.}~\bibnamefont
  {Desbuquois}}, \bibinfo {author} {\bibfnamefont {M.}~\bibnamefont {Messer}},
  \ and\ \bibinfo {author} {\bibfnamefont {T.}~\bibnamefont {Esslinger}},\
  }\href@noop {} {\bibfield  {journal} {\bibinfo  {journal} {Nat. Phys.}\
  }\textbf {\bibinfo {volume} {15}},\ \bibinfo {pages} {1161} (\bibinfo {year}
  {2019})}\BibitemShut {NoStop}%
\bibitem [{\citenamefont {Yao}\ \emph {et~al.}(2022)\citenamefont {Yao},
  \citenamefont {Zhang},\ and\ \citenamefont {Chin}}]{Yao2022}%
  \BibitemOpen
  \bibfield  {author} {\bibinfo {author} {\bibfnamefont {K.-X.}\ \bibnamefont
  {Yao}}, \bibinfo {author} {\bibfnamefont {Z.}~\bibnamefont {Zhang}}, \ and\
  \bibinfo {author} {\bibfnamefont {C.}~\bibnamefont {Chin}},\ }\href
  {https://www.nature.com/articles/s41586-021-04250-3} {\bibfield  {journal}
  {\bibinfo  {journal} {Nature}\ }\textbf {\bibinfo {volume} {602}},\ \bibinfo
  {pages} {68} (\bibinfo {year} {2022})}\BibitemShut {NoStop}%
\bibitem [{\citenamefont {Edmonds}\ \emph {et~al.}(2013)\citenamefont
  {Edmonds}, \citenamefont {Valiente}, \citenamefont
  {Juzeli\ifmmode~\bar{u}\else \={u}\fi{}nas}, \citenamefont {Santos},\ and\
  \citenamefont {\"Ohberg}}]{Edmonds2013}%
  \BibitemOpen
  \bibfield  {author} {\bibinfo {author} {\bibfnamefont {M.~J.}\ \bibnamefont
  {Edmonds}}, \bibinfo {author} {\bibfnamefont {M.}~\bibnamefont {Valiente}},
  \bibinfo {author} {\bibfnamefont {G.}~\bibnamefont
  {Juzeli\ifmmode~\bar{u}\else \={u}\fi{}nas}}, \bibinfo {author}
  {\bibfnamefont {L.}~\bibnamefont {Santos}}, \ and\ \bibinfo {author}
  {\bibfnamefont {P.}~\bibnamefont {\"Ohberg}},\ }\href {\doibase
  10.1103/PhysRevLett.110.085301} {\bibfield  {journal} {\bibinfo  {journal}
  {Phys. Rev. Lett.}\ }\textbf {\bibinfo {volume} {110}},\ \bibinfo {pages}
  {085301} (\bibinfo {year} {2013})}\BibitemShut {NoStop}%
\bibitem [{\citenamefont {Edmonds}\ \emph {et~al.}(2015)\citenamefont
  {Edmonds}, \citenamefont {Valiente},\ and\ \citenamefont
  {\"Ohberg}}]{Edmonds2015}%
  \BibitemOpen
  \bibfield  {author} {\bibinfo {author} {\bibfnamefont {M.~J.}\ \bibnamefont
  {Edmonds}}, \bibinfo {author} {\bibfnamefont {M.}~\bibnamefont {Valiente}}, \
  and\ \bibinfo {author} {\bibfnamefont {P.}~\bibnamefont {\"Ohberg}},\ }\href
  {\doibase 10.1209/0295-5075/110/36004} {\bibfield  {journal} {\bibinfo
  {journal} {Europhys. Lett.}\ }\textbf {\bibinfo {volume} {110}},\ \bibinfo
  {pages} {36004} (\bibinfo {year} {2015})}\BibitemShut {NoStop}%
\bibitem [{\citenamefont {Dingwall}\ and\ \citenamefont
  {\"Ohberg}(2019)}]{Dingwall2019}%
  \BibitemOpen
  \bibfield  {author} {\bibinfo {author} {\bibfnamefont {R.~J.}\ \bibnamefont
  {Dingwall}}\ and\ \bibinfo {author} {\bibfnamefont {P.}~\bibnamefont
  {\"Ohberg}},\ }\href {\doibase 10.1103/PhysRevA.99.023609} {\bibfield
  {journal} {\bibinfo  {journal} {Phys. Rev. A}\ }\textbf {\bibinfo {volume}
  {99}},\ \bibinfo {pages} {023609} (\bibinfo {year} {2019})}\BibitemShut
  {NoStop}%
\bibitem [{\citenamefont {Dingwall}\ \emph {et~al.}(2018)\citenamefont
  {Dingwall}, \citenamefont {Edmonds}, \citenamefont {Helm}, \citenamefont
  {Malomed},\ and\ \citenamefont {\"Ohberg}}]{Dingwall2018}%
  \BibitemOpen
  \bibfield  {author} {\bibinfo {author} {\bibfnamefont {R.~J.}\ \bibnamefont
  {Dingwall}}, \bibinfo {author} {\bibfnamefont {M.~J.}\ \bibnamefont
  {Edmonds}}, \bibinfo {author} {\bibfnamefont {J.~L.}\ \bibnamefont {Helm}},
  \bibinfo {author} {\bibfnamefont {B.~A.}\ \bibnamefont {Malomed}}, \ and\
  \bibinfo {author} {\bibfnamefont {P.}~\bibnamefont {\"Ohberg}},\ }\href
  {\doibase 10.1088/1367-2630/aab29e} {\bibfield  {journal} {\bibinfo
  {journal} {New J. Phys.}\ }\textbf {\bibinfo {volume} {20}},\ \bibinfo
  {pages} {043004} (\bibinfo {year} {2018})}\BibitemShut {NoStop}%
\bibitem [{\citenamefont {Bhat}\ \emph {et~al.}(2021)\citenamefont {Bhat},
  \citenamefont {Sivaprakasam},\ and\ \citenamefont {Malomed}}]{Bhat2021}%
  \BibitemOpen
  \bibfield  {author} {\bibinfo {author} {\bibfnamefont {I.~A.}\ \bibnamefont
  {Bhat}}, \bibinfo {author} {\bibfnamefont {S.}~\bibnamefont {Sivaprakasam}},
  \ and\ \bibinfo {author} {\bibfnamefont {B.~A.}\ \bibnamefont {Malomed}},\
  }\href {\doibase 10.1103/PhysRevE.103.032206} {\bibfield  {journal} {\bibinfo
   {journal} {Phys. Rev. E}\ }\textbf {\bibinfo {volume} {103}},\ \bibinfo
  {pages} {032206} (\bibinfo {year} {2021})}\BibitemShut {NoStop}%
\bibitem [{\citenamefont {Sanz}\ \emph {et~al.}(2022)\citenamefont {Sanz},
  \citenamefont {Fr\"olian}, \citenamefont {Chisholm}, \citenamefont
  {Cabrera},\ and\ \citenamefont {Tarruell}}]{Sanz2022}%
  \BibitemOpen
  \bibfield  {author} {\bibinfo {author} {\bibfnamefont {J.}~\bibnamefont
  {Sanz}}, \bibinfo {author} {\bibfnamefont {A.}~\bibnamefont {Fr\"olian}},
  \bibinfo {author} {\bibfnamefont {C.~S.}\ \bibnamefont {Chisholm}}, \bibinfo
  {author} {\bibfnamefont {C.~R.}\ \bibnamefont {Cabrera}}, \ and\ \bibinfo
  {author} {\bibfnamefont {L.}~\bibnamefont {Tarruell}},\ }\href {\doibase
  10.1103/PhysRevLett.128.013201} {\bibfield  {journal} {\bibinfo  {journal}
  {Phys. Rev. Lett.}\ }\textbf {\bibinfo {volume} {128}},\ \bibinfo {pages}
  {013201} (\bibinfo {year} {2022})}\BibitemShut {NoStop}%
\bibitem [{\citenamefont {Snyder}\ and\ \citenamefont
  {Mitchell}(1997)}]{Snyder1997}%
  \BibitemOpen
  \bibfield  {author} {\bibinfo {author} {\bibfnamefont {A.~W.}\ \bibnamefont
  {Snyder}}\ and\ \bibinfo {author} {\bibfnamefont {D.~J.}\ \bibnamefont
  {Mitchell}},\ }\href@noop {} {\bibfield  {journal} {\bibinfo  {journal}
  {Science}\ }\textbf {\bibinfo {volume} {276}},\ \bibinfo {pages} {1538}
  (\bibinfo {year} {1997})}\BibitemShut {NoStop}%
\bibitem [{\citenamefont {Pitaevskii}\ and\ \citenamefont
  {Stringari}(1999)}]{Pitaevskii1999}%
  \BibitemOpen
  \bibfield  {author} {\bibinfo {author} {\bibfnamefont {L.}~\bibnamefont
  {Pitaevskii}}\ and\ \bibinfo {author} {\bibfnamefont {S.}~\bibnamefont
  {Stringari}},\ }\href
  {https://journals.aps.org/prl/abstract/10.1103/PhysRevLett.83.4237}
  {\bibfield  {journal} {\bibinfo  {journal} {Phys. Rev. Lett.}\ }\textbf
  {\bibinfo {volume} {83}},\ \bibinfo {pages} {4237} (\bibinfo {year}
  {1999})}\BibitemShut {NoStop}%
\bibitem [{\citenamefont {Satsuma}\ and\ \citenamefont
  {Yajima}(1974)}]{Satsuma1974}%
  \BibitemOpen
  \bibfield  {author} {\bibinfo {author} {\bibfnamefont {J.}~\bibnamefont
  {Satsuma}}\ and\ \bibinfo {author} {\bibfnamefont {N.}~\bibnamefont
  {Yajima}},\ }\href@noop {} {\bibfield  {journal} {\bibinfo  {journal} {Prog.
  Theor. Phys. Supp.}\ }\textbf {\bibinfo {volume} {55}},\ \bibinfo {pages}
  {284} (\bibinfo {year} {1974})}\BibitemShut {NoStop}%
\bibitem [{\citenamefont {Jackiw}(1997)}]{Jackiw1997}%
  \BibitemOpen
  \bibfield  {author} {\bibinfo {author} {\bibfnamefont {R.}~\bibnamefont
  {Jackiw}},\ }\href@noop {} {\bibfield  {journal} {\bibinfo  {journal}
  {Nonlinear Math. Phys.}\ }\textbf {\bibinfo {volume} {4}},\ \bibinfo {pages}
  {261} (\bibinfo {year} {1997})}\BibitemShut {NoStop}%
\end{thebibliography}%

\end{document}